\documentclass[11pt]{article}
\pdfoutput=1

\usepackage{jheppub}
\usepackage{url,comment}
\usepackage{times}
\usepackage{latexsym}
\usepackage{graphicx, graphics, hyperref, amsmath, amssymb, slashed, xcolor, bbm,bm,amsthm, array}
 \usepackage{subfigure}
 \usepackage{listings} 
 \usepackage{trimclip}

\usepackage{rotating}
\usepackage{afterpage}
 \usepackage[colorlinks=true,citecolor=blue,urlcolor=blue]{hyperref}

\newcommand{\nc}{\newcommand}

\nc{\beq}{\begin{equation}}
\nc{\eeq}{\end{equation}}
\nc{\barray}{\begin{eqnarray}}
\nc{\earray}{\end{eqnarray}}
\nc{\barrayn}{\begin{eqnarray*}}
\nc{\earrayn}{\end{eqnarray*}}
\nc{\bcenter}{\begin{center}}
\nc{\ecenter}{\end{center}}
\nc{\mc}{\mathcal}
\nc{\er}[1]{(\ref{eq:#1})}
\nc{\onehalf}{\frac{1}{2}} 
\nc{\partialbar}{\bar{\partial}}
\nc{\psit}{\widetilde{\psi}}
\nc{\Tr}{\mbox{Tr}}
\nc{\hc}{\mbox{H.c.}}
\nc{\ev}{\;\mathrm{eV}}
\nc{\mev}{\;\mathrm{MeV}}
\nc{\gev}{\;\mathrm{GeV}}
\nc{\kev}{\;\mathrm{keV}}
\nc{\tev}{\;\mathrm{TeV}}

\def\chii0{\chi_i^0}
\def\chij0{\chi_j^0}

\newcommand{\gsim}{\lower.7ex\hbox{$\;\stackrel{\textstyle>}{\sim}\;$}}
\newcommand{\lsim}{\lower.7ex\hbox{$\;\stackrel{\textstyle<}{\sim}\;$}}
\nc{\ttbar}{t\bar t}

\def\dbar{{\mathchar'26\mkern-12mu d}}

\def\beq{\begin{equation}}
\def\eeq{\end{equation}}
\def\bea{\begin{eqnarray}}
\def\eea{\end{eqnarray}}

\newcommand{\cref}[1]{Chapter~\ref{c.#1}}

\def\beq{\begin{equation}}
\def\eeq{\end{equation}}
\def\bea{\begin{eqnarray}}
\def\eea{\end{eqnarray}}

\graphicspath{{plots/}}

\newcommand{\Planck}{{\sc {\it Planck}}}

\preprint{YITP-SB-2021-27}

\title{Improved cosmological constraints on the neutrino mass and lifetime}

\author[a]{Guillermo Franco Abellán,}

\author[b]{Zackaria Chacko,}

\author[c]{Abhish Dev,}

\author[d]{Peizhi Du,}

\author[a]{Vivian Poulin,}

\author[e]{and Yuhsin Tsai}

\affiliation[a]{Laboratoire Univers \& Particules de Montpellier (LUPM), CNRS \& Universit\'e de Montpellier (UMR-5299), Place Eug\`ene Bataillon, F-34095 Montpellier Cedex 05, France}

\affiliation[b]{Maryland Center for Fundamental Physics, Department of Physics, University of Maryland, College Park, MD 20742-4111 USA}

\affiliation[c]{Theoretical Physics Department, Fermilab, P.O. Box 500, Batavia, IL 60510, USA}

\affiliation[d]{C.N. Yang Institute for Theoretical Physics, Stony Brook University, Stony Brook, NY, 11794, USA}

\affiliation[e]{Department of Physics, University of Notre Dame, IN 46556, USA}

\date{\today}

\abstract{
We present cosmological constraints on the sum of neutrino masses as a function
of the neutrino lifetime, in a framework in which neutrinos decay into dark radiation after becoming non-relativistic.  We find that in this regime the cosmic microwave background (CMB), baryonic acoustic oscillations (BAO) and (uncalibrated) luminosity distance to supernovae from the Pantheon catalog constrain the sum of neutrino masses $\sum m_\nu$ to obey  $\sum m_\nu< 0.42$ eV at (95$\%$ C.L.). While the bound has improved significantly as compared to the limits on the same scenario from {\it Planck} 2015, it still represents a significant relaxation of the constraints as compared to the stable neutrino case.
We show that most of the improvement can be traced to the more precise measurements of low-$\ell$ polarization data in {\it Planck} 2018, which leads to tighter constraints on  $\tau_{\rm reio}$ (and thereby on $A_s$), breaking the degeneracy arising from the effect of (large) neutrino masses on the amplitude of the CMB power spectrum. 
}

\begin{document}

\maketitle

\section{Introduction}

Even though neutrinos were first detected more than six decades ago, they remain among the most mysterious particles in nature, with many of their fundamental properties still to be determined. In particular, although oscillation experiments have provided convincing evidence that neutrinos have non-vanishing masses, these measurements are only sensitive to the mass-squared splittings and consequently the spectrum of neutrino masses remains unknown. The lifetimes of the neutrinos are also poorly constrained, especially in comparison to the other particles in the Standard Model (SM). The determination of the masses and the lifetimes of these mysterious particles remain some of the most important open problems in fundamental physics.  

The fact that cosmic neutrinos are among the most abundant particles in the universe, contributing significantly to the total energy density at early times, provides an opportunity to measure their properties. In particular, the evolution of the cosmological density fluctuations depends on $\sum m_\nu$, the sum of neutrino masses. This translates into characteristic effects on the cosmic microwave background (CMB) and large-scale structure (LSS)~\cite{Bond:1980ha,Hu:1997mj} (for reviews see~\cite{Wong:2011ip,Lesgourgues:2018ncw,Tanabashi:2636832,Lattanzi:2017ubx}), that are large enough to allow the sum of neutrino masses to be determined in the near future. This determination is based on the observation that massive neutrinos contribute differently to cosmological observables than either massless neutrinos or cold dark matter (CDM). At early times, while still relativistic, massive neutrinos contribute to the energy density in radiation, just as in the case of massless neutrinos. However, after neutrinos become non-relativistic, their energy density redshifts as matter and therefore contributes more to the expansion rate than massless neutrinos, which would continue to redshift as radiation. As a result, over a given redshift span, the higher expansion rate reduces the time available for the growth of matter density perturbations. However, since massive neutrinos retain pressure until late times, their contribution to the density perturbations on scales  below their free streaming lengths is too small to compensate for the shorter structure formation  time. Therefore, if neutrinos become non-relativistic after recombination, the net effect of non-vanishing neutrino masses is a suppression of the matter power spectrum and the CMB lensing potential. Based on this, current observations are able to place a bound on the sum of neutrino masses, $\sum m_\nu \lesssim 0.12$ eV~\cite{Aghanim:2018eyx}. It is important to note that this result assumes that neutrinos are stable on timescales of order the age of the universe. In scenarios in which the neutrinos decay~\cite{Serpico:2007pt,Serpico:2008zza}, or annihilate away into lighter species~\cite{Beacom:2004yd,Farzan:2015pca} on timescales shorter than the age of the universe, this bound is no longer valid and must be reconsidered. 

Cosmological observations can also be used to place limits on the neutrino lifetime. In the case of neutrinos that decay to final states containing photons, the bounds on spectral distortions in the cosmic microwave background (CMB) can be translated into limits on the neutrino lifetime, $\tau_\nu \gtrsim 10^{19}$ s for the larger mass splitting and $\tau_\nu \gtrsim 4 \times 10^{21}$ s for the smaller one~\cite{Aalberts:2018obr}. In the case of decays to invisible final states, the limits are much weaker. For neutrinos that decay while still relativistic, the decay and inverse decay processes can prevent neutrinos from free streaming. 
Measurements of the CMB power spectra set a lower bound on the neutrino lifetime, $\tau_{\nu} \geq 4 \times 10^6$ s $(m_\nu /0.05 \textrm{eV})^5$, in the case of decay into dark radiation~\cite{Barenboim:2020vrr} (for earlier work see
\cite{Hannestad:2005ex,Basboll:2008fx,Archidiacono:2013dua,Escudero:2019gfk}). 
In the case of non-relativistic neutrino decays into dark radiation, the energy density of the decay products redshifts faster than that of stable massive neutrinos. Unstable neutrinos therefore have less of an effect on structure formation than stable neutrinos of the same mass. 
Consequently, cosmological observables depend both on the masses of the neutrinos and their lifetimes, and heavier values of  $\sum m_\nu$ may still be allowed by the data provided the neutrino lifetime is short enough. In Ref.~\cite{Chacko:2019nej}, \Planck~2015 and LSS data were used to place constraints on the neutrino mass as a function of the lifetime, and found that values of $\sum m_{\nu}$ as large as $0.9$ eV were still allowed by the data. Future LSS measurements at higher redshifts may be able to break the degeneracy between the neutrino mass and lifetime and measure these parameters independently~\cite{Chacko:2020hmh}. It is worth noting that there are also bounds on the neutrino lifetime from Supernova~1987A~~\cite{Frieman:1987as}, solar neutrinos~\cite{Joshipura:2002fb,Beacom:2002cb,Bandyopadhyay:2002qg,Berryman:2014qha}, astrophysical neutrinos measured at IceCube~\cite{Baerwald:2012kc,Pagliaroli:2015rca,Bustamante:2016ciw,Denton:2018aml,Abdullahi:2020rge,Bustamante:2020niz}, atmospheric neutrinos and long baseline experiments~\cite{GonzalezGarcia:2008ru,Gomes:2014yua,Choubey:2018cfz,Aharmim:2018fme}. However,these constraints are in general much weaker than the limits from cosmology.

In this paper we revisit the scenario in which neutrinos decay into dark radiation after becoming non-relativistic and obtain updated limits based on the newer data from \Planck~2018. In order to take advantage of the greater precision of the new data, the analysis we perform is also more accurate. We find that, under the assumption that neutrinos decay after becoming non-relativistic, the neutrino mass bound from \Planck~2018 data (in combination with BOSS baryon acoustic oscillation (BAO) data and Pantheon SN1a data) is relaxed to $\sum m_\nu \lesssim 0.42 $ eV (95\% C.L.).\footnote{It is a factor of two weaker than the constraints advocated in Ref.~\cite{Lorenz:2021alz}, which used a model-independent approach to constrain the neutrino mass as a function of redshift, but neglected the effect of the daughter particles.} 
While  this  represents  a remarkable  relaxation  of  the  constraints as compared to the case of stable neutrinos, we note  that it is much stronger than the limit derived from \Planck~2015 data for the same decaying neutrino scenario, $\sum m_\nu \lesssim 0.9$ eV at (95\% C.L.). 
We show that the improvement of the bound arises primarily from the more precise low-$\ell$ polarization data from \Planck~2018, which allows an improved determination of the optical depth to reionization $\tau_{\rm reio}$, thereby breaking the correlation with $\sum m_\nu$ that appears (for relatively high neutrino masses) through the impact of neutrinos on the overall height of the acoustic peaks (i.e.  the ``early integrated Sachs-Wolfe effect")~\cite{Lesgourgues:2018ncw}.

Besides using up-to-date cosmological data, we also improve the analysis from Ref.~\cite{Chacko:2019nej} by incorporating higher order corrections due to neutrino decays into the Boltzmann equations that describe the evolution of Universe's energy and metric fluctuations. Recently, Ref.~\cite{Barenboim:2020vrr} provided a complete set of Boltzmann equations for the neutrino decay, but did not conduct Markov Chain Monte Carlo (MCMC) runs necessary to calculate updated neutrino bounds. In this work, we derive Boltzmann equations exactly valid in the absence of `inverse-decays' and quantum statistics. For the numerical implementation, we follow a consistent $T_{\rm dec}/m_\nu$ expansion, where $T_{\rm dec}$ is the temperature at the time of the decay, so that the analysis is under control when neutrinos decay after become non-relativistic.

This paper is organized as follows. In section~\ref{sec:param_space}, we present a summary of constraints on the parameter space of decaying neutrinos. In section~\ref{sec:boltzmann_equations}, we derive the set of Boltzmann equations to describe neutrino decay that are valid in the non-relativistic regime and compare our improved analysis to past work. In section~\ref{sec:MCstudy}, we present a MCMC analysis of the decaying neutrino scenario against up-to-date cosmological data. Finally, we conclude in section~\ref{sec:conclusions}.

\section{Parameter space of decaying neutrinos}\label{sec:param_space}
\begin{figure}
\begin{center}
\includegraphics[scale=0.7]{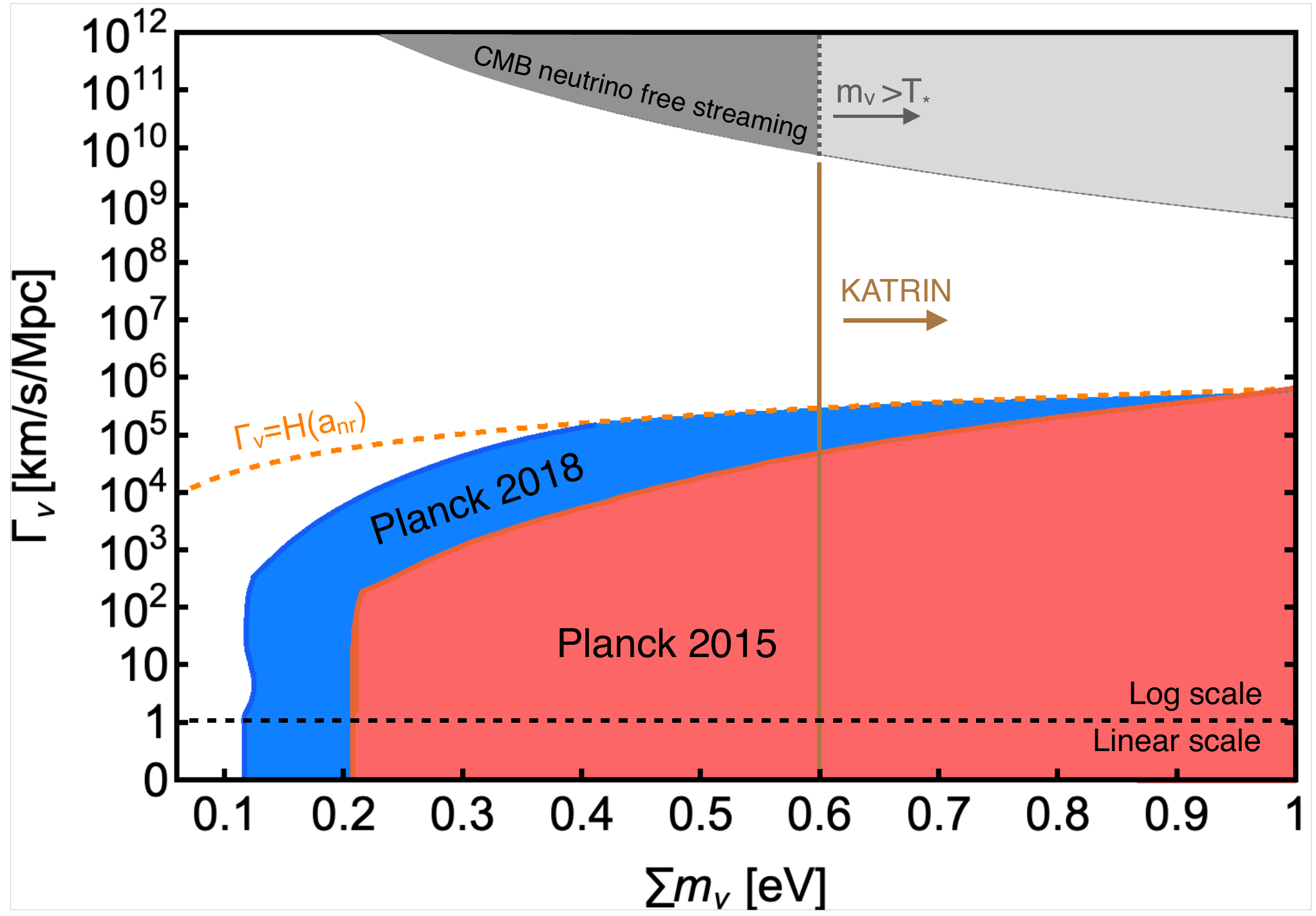}
 \caption{The plot shows the current constraints on decaying neutrinos in the $\sum m_\nu-\Gamma_\nu$ parameter space. The colored regions are excluded by 
current data while the white region is allowed. The orange dashed line 
represents $\Gamma_\nu=H(a_{\rm nr})$. Our study focuses on the region below this line, 
meaning decay happens after neutrinos have become non-relativistic. The grey region shows 
current constraints on neutrino mass and lifetime coming from the requirement that neutrinos are free streaming close to recombination~\cite{Barenboim:2020vrr}. The light grey region indicates that this bound may not be applicable when neutrino mass is larger than the temperature of recombination: $m_\nu>T_*\sim 0.2$ eV~\cite{Barenboim:2020vrr}. Our analysis excludes the red (blue) region labelled ``Planck 2015''(``Planck 2018'') based on the data (Planck+BAO+Pantheon). The vertical brown line shows the projected KATRIN sensitivity. }
 \label{TargetSpace}
 \end{center}
 \end{figure}

In this section we outline the constraints on the mass and lifetime of neutrinos decaying into
dark radiation. As explained in the introduction, current cosmological observables only 
place limits on a combination of the sum of neutrino masses and their lifetime. 
Therefore, in this study we will map out the constraints 
in the two-dimensional parameter space spanned by the sum of neutrino 
masses ($\sum m_\nu$) and the neutrino decay width ($\Gamma_\nu$), as 
shown in Fig.~\ref{TargetSpace}. 
In our analysis we assume that all three neutrinos are degenerate in mass. 
This is a good approximation because the current bounds on $\sum m_\nu$ are larger than 
the observed mass splittings (see Fig.~\ref{TargetSpace}). 
We further assume that all three 
neutrinos have the same decay width $\Gamma_{\nu}$. 
Since the mixing 
angles in the neutrino sector are large, this is a good approximation in 
many simple models of decaying neutrinos if the spectrum of neutrinos is 
quasi-degenerate. 
While this is a simple parameterization of neutrino decays, our bounds can easily be applied to specific models, as done in great details in Ref.~\cite{Escudero:2020ped}.

The CMB can be used to constrain the masses and decay widths of neutrinos that decay prior to recombination.When neutrinos decay while still relativistic, decay and inverse decay can prevent neutrinos from free-streaming. 
If this happens before recombination,  it can alter the well-known `neutrino drag' effect that manifests as a phase-shift at high-$\ell$'s in the CMB power spectrum \cite{Bashinsky:2003tk,Audren:2014lsa,Follin:2015hya,Baumann:2015rya}.
Therefore, CMB data can place a constraint on the decay width of neutrinos. 
 The resulting bound depends on neutrino masses, and was recently updated in Ref.~\cite{Barenboim:2020vrr},
$\tau_\nu \geq 4 \times 10^{6}\,\textrm{s} 
\left({m_\nu}/{0.05\,\textrm{eV}}\right)^5$~. 
This bound excludes the grey region at the top of 
Fig.~\ref{TargetSpace}. 

 In addition, based on the analysis in this paper, part of the `late-decay' parameter space can also be excluded based on the gravitational impacts of massive neutrinos on the CMB and LSS. Through the Monte Carlo study 
presented in section~\ref{sec:MCstudy}, the blue (red) shaded region in Fig.~\ref{TargetSpace} is excluded by the data combination \Planck~2018(2015)+BAO+Pantheon.\footnote{Note that in our analysis we scanned the region between 
$0\leq{\rm log}_{10}\frac{\Gamma_\nu}{\textrm{km/s/Mpc}}\leq6$. 
In 
Fig.~\ref{TargetSpace}, we have extrapolated the bound at ${\rm 
log}_{10}\frac{\Gamma_\nu}{\textrm{km/s/Mpc}}=0$ to $\Gamma_\nu=0$, 
because the constraint on $\sum m_\nu$ is independent of $\Gamma_\nu$ 
when $\Gamma_\nu\ll H_0$.}
The orange dashed line in the figure ($\Gamma_\nu=H(a_{\rm nr})$) separates the region 
where neutrinos decay when non-relativistic from the region where they 
decay while still relativistic. 
Here $a_{\rm nr}$ corresponds to the approximate 
scale factor at the time that neutrinos transition to non-relativistic, and is defined as $3 T_\nu(a_{\rm nr})=m_\nu$.  
This simple
definition is based on the fact that for relativistic neutrinos at 
temperature $T_\nu$, the average energy per neutrino is approximately $3 
T_\nu$. 
The Hubble scale at $a_{\rm nr}$ is given by, 
  \begin{eqnarray}
      H(a_{\rm nr}) &=& H_0\sqrt{\Omega_m}\bigg(\frac{\sum m_\nu}{9T_{\nu0}}\bigg)^{3/2} 
\\
&\simeq&  7.5\times 10^5 {\rm km/s/Mpc} \bigg(\frac{H_0}{68 {\rm km/s/Mpc}}\bigg)
\bigg(\frac{\Omega_m}{0.3}\bigg)^{1/2}\bigg(\frac{\sum m_\nu}{1 {\rm eV}}\bigg)^{3/2}\bigg(\frac{1.5\times 10^{-4} {\rm eV}}{T_{\nu0}}\bigg)^{3/2},\nonumber
  \end{eqnarray}
where $T_{\nu0}$ is the present neutrino temperature.
Since our study focuses on the decay of neutrinos after they become
non-relativistic, we only present constraints below the orange 
dashed line. Our analysis shows that $\sum m_\nu$ as large as $0.42$ eV is still allowed by the data.

Our results have important implications for current and future laboratory experiments designed to detect neutrino masses. 
Next generation tritium decay experiments such as KATRIN~\cite{Angrik:2005ep} are expected to be sensitive to values of $m_{\nu_e}$ as low as $0.2$ eV, corresponding to $\sum m_{\nu}$ of order 0.6 eV. 
Naively, a signal in these experiments would conflict with the current cosmological bound for stable neutrinos, $\sum m_{\nu} < 0.12$ eV. However, since the unstable neutrino paradigm greatly expands the range of neutrino masses allowed by current cosmological data, it is interesting to explore whether this scenario can accommodate a potential signal at KATRIN.
In Fig.~\ref{TargetSpace}, we display a brown vertical line $\sum m_{\nu} 
= 0.6$ eV that corresponds to the expected KATRIN sensitivity. We see
that this value of $\sum m_\nu$ is too large to be accommodated in the non-relativistic decay regime, where our analysis is valid. 
However, our result, in combination with those from the `relativistic decay' scenario studied in Ref.~\cite{Barenboim:2020vrr}, leaves open the interesting possibility that neutrinos decaying with a decay width between $\log_{10}\frac{\Gamma_\nu}{\rm km/s/Mpc} \sim 5.5-9$ could reconcile 
cosmological observations with a potential detection at KATRIN, thereby opening a large discovery potential for laboratory experiments.
To confirm this conjecture, more work needs to be done to cover the `intermediate' decay regime (i.e. where neutrinos are neither fully relativistic nor fully non-relativistic). We leave this for future work.

{In recent years, a number of studies have attempted to constrain the neutrino mass ordering, showing that under the assumption of stable neutrinos, the inverted ordering is now disfavored by constraints from joint analysis of cosmological and oscillation data \cite{Gerbino:2016ehw,Caldwell:2017mqu,Vagnozzi:2017ovm,Simpson:2017qvj,DiValentino:2021hoh,Jimenez:2022dkn} (see also Refs.~\cite{Schwetz:2017fey,Gariazzo:2018pei,Hergt:2021qlh,Gariazzo:2022ahe} for a different take) as well as from Ly-$\alpha$ observations \cite{Palanque-Delabrouille:2019iyz}. However, these arguments are centered on the fact that these analysis lead to a constraint on $\sum m_\nu$ at odds with the lower bound on the sum of neutrino masses in the case of inverted ordering, $\sum m_\nu\gtrsim 0.1$ eV. Our result suggests that these constraints are strongly dependent on the assumption of neutrino stability over cosmological timescales, and therefore that the inverted ordering is not robustly excluded. 
It would be very interesting to extend our analysis to the inclusion of Ly-$\alpha$ data to confirm this conclusion.}

\section{Boltzmann equations for massive neutrinos decaying into radiation }

\label{sec:boltzmann_equations}

In this section, we revisit the set of Boltzmann equations describing the evolution of the phase space distribution (PSD) of massive particles decaying into daughter radiation. In our analysis, we assume the decay happens after the neutrinos have become non-relativistic so that the contribution from inverse decay processes can be safely neglected.

\subsection{Derivation of the equations}
We denote the phase space distribution of each species as $f(q,\hat n,\vec{x},\tau)$, which is a function of the comoving momentum $q \hat n$, coordinates $\vec{x}$ and conformal time $\tau$. The general time evolution of $f$ is controlled by the Boltzmann equations, 
\begin{eqnarray}
\frac{df}{d\tau}=\frac{\partial f}{\partial \tau}+\frac{d x^i}{d\tau}\frac{\partial f}{\partial x^i}+\frac{d q}{d\tau}\frac{\partial f}{\partial q}+\frac{d \hat n}{d\tau}\cdot\frac{\partial f}{\partial \hat n}=C[f],
\end{eqnarray}
where $C[f]$ is the collision term that includes all the processes involving the species.

This phase space distribution  has the leading order contribution $\bar f(q,\tau)$ that only depends on $q$ and $\tau$, while perturbations are encoded in $\Delta f(q,\hat n,\vec{x},\tau)$,
\begin{eqnarray}
f(q,\hat n,\vec{x},\tau)\equiv \bar f(q,\tau)+\Delta f(q,\hat n,\vec{x},\tau).
\end{eqnarray}

Treating $\Delta f$ fluctuations about the 
homogeneous background as higher order perturbations, the zeroth order
Boltzmann equations for $\bar f$ take the form
 \begin{eqnarray}\label{eq:BK_general} 
\frac{\partial \bar f}{\partial \tau}= C[\bar f].
 \end{eqnarray} 
 In this work, our focus is on the case in which neutrinos decay after turning non-relativistic. In this scenario, we can neglect the effects of inverse decay processes and quantum statistics. The collision term for the neutrino and its daughters are respectively given by~\cite{Chacko:2019nej}
\begin{eqnarray}
C_\nu=&-&\frac{a^2}{2\epsilon_{\nu}}\int \prod_i\frac{\dbar^3 q_i}{2\epsilon_{i}}|\mathcal{M}|^2(2\pi)^4 \delta^{(4)}(q-\Sigma_i q_{i})f_\nu(q),\\
C_{Dj}=&+&\frac{a^2}{2\epsilon_{j}}\int \frac{\dbar^3 q}{2\epsilon_\nu}\prod_{i\neq j}\frac{\dbar q_i^3}{2\epsilon_{i}}|\mathcal{M}|^2(2\pi)^4 \delta^{(4)}(q-\Sigma_i q_{i}) f_\nu(q).\label{eq:C_D}
 \end{eqnarray}
 Here  $\dbar^3q\equiv d^3 q/(2\pi)^3 $, $\epsilon\equiv\sqrt{q^2+a^2m^2}$ represents the comoving energy and $a$ is the scale factor. The label $i(j)$ denotes the $i$th($j$th) daughter.
In the case of two body decays to massless daughters, the amplitude squared $|\mathcal{M}|^2$ is simply related to the rest-frame decay width of the neutrino as $|\mathcal{M}|^2=16 \pi \Gamma_\nu m_\nu.$ 
From the collision terms above, the background evolution for decaying neutrinos is given by
  \begin{eqnarray}\label{eq:BK_nu} 
\frac{\partial \bar f_\nu}{\partial \tau}&=& -a\frac{\Gamma_\nu}{\gamma}\bar f_\nu,
 \end{eqnarray} 
 where $\Gamma_\nu$ is the neutrino decay width and $\gamma$ is the Lorentz boost factor,
 \begin{equation}
 \gamma\!=\!\frac{\sqrt{q^2\!+\!a^2 m_\nu^2}}{(a m_\nu)}.
 \end{equation}
The formal solution to $\bar f_\nu(q,\tau)$ from the differential equation Eq.~(\ref{eq:BK_nu})  is 
 \begin{eqnarray}
 \label{eq:formal}
\bar f_\nu(q,\tau)=\bar f_{\rm ini}(q)e^{-\Gamma_\nu \int_{\tau_{\rm ini}}^\tau \frac{ 
a}{\gamma(a)} d\tau'}, 
 \end{eqnarray} 
 where $\tau_{\rm ini}$ denotes the initial conformal time and $\bar f_{\rm ini}(q)$ 
represents the initial momentum distribution, which we take to be of the Fermi-Dirac form, $\bar f_{\rm ini}=1/(e^{q/T_{\nu0}}+1)$. 

The Boltzmann equations for the individual daughter particles do not have a simple form, especially when the daughters consist of more than two species. However, since the daughter particles are taken to be massless in this study the total background density of daughter radiation can be defined as
\begin{eqnarray}\label{eq:rho_D_general}
\bar\rho_D\equiv 4\pi a^{-4}\sum_i \int dq\, q^3 \bar f_{Di}(q),
\end{eqnarray}
where $\bar f_{Di}$ is the background phase space distribution of the $i$th daughter particle. With the definition in Eq.~(\ref{eq:rho_D_general}), regardless of the number of daughter particles and their spins, the Boltzmann equation for the total background daughter density $\bar \rho_D$ has the simple form  
 \begin{equation}
\label{IBD} 
\frac{\partial \bar{\rho}_{D}}{\partial\tau} + 4 a H \bar{\rho}_{D} = a \Gamma_\nu m_\nu \bar{n}_{\nu} ,
 \end{equation}
 where $\bar n_\nu\equiv 4\pi a^{-3}\int dq\, q^2 \bar f_{\nu}( q)$.
 
 We now turn to the Boltzmann equations describing the perturbations of the phase space distribution of decaying neutrinos and their decay products. We work in the synchronous gauge for which the metric perturbations can be parametrized as \cite{Ma:1994dv}
 \begin{eqnarray}
 ds^2=a^2[-d\tau^2+(\delta_{ij}+H_{ij})dx^idx^j].
 \end{eqnarray}
In Fourier space, $H_{ij}$ is given by
\begin{eqnarray}
H_{ij}(\vec k ,\tau)=\hat k_i\hat k_j h(\vec k,\tau)+\left(\hat k_i\hat k_j -\frac{1}{3}\delta_{ij}\right)6\eta(\vec k,\tau),
\end{eqnarray}
where $\vec k$ is conjugate to $\vec x$ and $h$ and $\eta$ are the two independent scalar metric perturbations. To obtain the Boltzmann hierarchy, we expand the angular dependence of the perturbations as a series in Legendre
polynomials,
\begin{eqnarray}\label{LegExp}
\Delta f(q,\hat n,\vec{k},\tau)=\sum_{\ell=0}^{\infty}(-i)^\ell(2\ell+1)\Delta f_\ell(q,k,\tau)P_\ell (\hat k\cdot \hat n),
\end{eqnarray}
where $P_\ell$ represents the $\ell$th Legendre polynomial.
The Boltzmann hierarchy for the perturbations of the decaying massive neutrinos $\Delta f_{\nu(\ell)}$ read \cite{Chacko:2019nej}
\begin{align}
&\Delta \dot{f}_{\nu(0)}  = -\frac{q k}{\varepsilon_\nu} \Delta f_{\nu(1)} 
 + q\frac{\partial \bar f_\nu}{\partial q}  \frac{\dot{h}}{6} - \frac{a^2 \Gamma_\nu m_\nu}{\varepsilon_\nu } \Delta f_{\nu(0)}
, \label{delta_f_nu_0}\\ 
&\Delta \dot{f}_{\nu (1)}  = \frac{q k}{3\varepsilon_\nu  } \left[\Delta f_{\nu(0)}-2 \Delta f_{\nu(2)} \right]- \frac{a^2 \Gamma_\nu m_\nu}{\varepsilon_\nu } \Delta f_{\nu(1)}
, \label{delta_f_nu_1}\\ 
&\Delta \dot{f}_{\nu (2)}  = \frac{q k}{5\varepsilon_\nu } \left[ 2 \Delta f_{\nu(1)}-3 \Delta f_{\nu(3)} \right] 
 - q\frac{\partial \bar f_\nu}{\partial q} \frac{(\dot{h}+6\dot{\eta})}{15}  -\frac{a^2 \Gamma_\nu m_\nu}{\varepsilon_\nu } \Delta f_{\nu(2)},   \  \label{delta_f_nu_2} \\ 
&\Delta \dot{f}_{\nu(\ell>2)}  = \frac{q k}{(2\ell+1)\varepsilon_\nu } \left[ \ell \Delta f_{\nu(\ell-1)}-(\ell+1) \Delta f_{\nu(\ell+1)} \right] -\frac{a^2 \Gamma_\nu m_\nu}{\varepsilon_\nu } \Delta f_{\nu(\ell)}.  \label{delta_f_nu_3}
\end{align}
Here $\varepsilon_\nu = \sqrt{q^2 + a^2 m_\nu^2}$ indicates the comoving energy of the neutrinos. 

To study the perturbations of the daughter radiation, we focus on the case of two-body decay. In this case, the phase space distributions of the two massless particles are basically identical and they can be considered effectively as one species with a single $f_D$. We can therefore define multipoles $F_{D(\ell)}$ as in Ref.~\cite{Poulin:2016nat},
\begin{eqnarray}
F_{D(\ell)}\equiv \frac{4\pi}{\rho_c} \int dq q^3 \Delta f_{\rm D(\ell)},
\end{eqnarray}
where $\rho_c$ is the critical density of Universe today.
The Boltzmann hierarchy of the $F_{D(\ell)}$ can be written as, 
\begin{align}
 &\dot{F}_{D(0)}   = - k F_{D(1)}
 -\frac{2}{3} \dot{h} \,r_{D}+ C_0,  \nonumber  \\ 
 & \dot{F}_{D(1)}  = \frac{ k}{3} F_{D(0)}-\frac{2k}{3} F_{D(2)} +C_1, \nonumber\\ 
 &\dot{F}_{D(2)}  = \frac{2 k}{5} F_{D(1)}-\frac{3 k}{5} F_{D(3)}
 + \frac{4 (\dot{h}+6\dot{\eta})}{15}  r_{D} +C_2,\nonumber \\
 &\dot{F}_{D(\ell>2)} = \frac{ k}{(2\ell+1)} \left[ \ell F_{D(\ell-1)}-(\ell+1) F_{D(\ell+1)} \right]+C_\ell, \label{eq:F_r}
\end{align}
where  $r_{D} \equiv a^4\bar{\rho}_{D}/\rho_c$. The terms $C_{\ell}$ appearing in Eq.  (\ref{eq:F_r}) arise from the integrated daughter collision term in Eq.~(\ref{eq:C_D}) expanded in terms of Legendre polynomials. The expression for $C_{\ell}$ is given by,
\begin{align}\label{eq:CLs}
C_{\ell} &= 2 i^\ell\int\frac{d\Omega_k}{4\pi} P_{\ell}(\hat{q}_1\cdot\hat{k})\left( \frac{4\pi}{\rho_c}\int dq_1 q_1^3 C_{D1}[q_1,\hat{q}_1\cdot\hat{k}]\right),\notag\\
&=  i^\ell\left(\frac{32 \pi m_\nu \Gamma_\nu a^2}{\rho_c}\right)\int d\Omega_k P_{\ell}(\hat{q}_1\cdot\hat{k})\int \frac{dq_1}{2\epsilon_1} q_1^3\int \frac{\dbar^3 q_2}{2\epsilon_2}\frac{\dbar^3 q}{2\epsilon_\nu} \Delta f_\nu(q,\hat{q}\cdot\hat{k}) (2\pi)^4\delta^{(4)}(q -q_1-q_2).
\end{align}
 The overall factor of two in the equation above arises because we are adding the collision integrals of the two massless daughters, which are of the same form. In this expression $d\Omega_k$ represents the differential solid angle along the direction $\hat{k}$, while $q_{1,2}$ are the momenta of daughter particles. The $\dbar^3 q_2$ integral can be easily evaluated using the delta function corresponding to momentum conservation. In order to perform the integral over $d\Omega_k$, we notice that the direction of $\hat{k}$ enters only via  $P_{\ell}(\hat{q}_1\cdot\hat{k})$ and $\Delta f_\nu(q,\hat{q}\cdot\hat{k})$. Now, using the Legendre expansion of $\Delta f_\nu(q,\hat{q}\cdot\hat{k})$ in Eq. (\ref{LegExp}) and employing the identity
\begin{equation}
   \int d\Omega_k P_{\ell}(\hat{k}\cdot\hat{q})P_{\ell'}(\hat{k}\cdot\hat{q}_1) = \left(\frac{4\pi}{2\ell + 1}\right)P_{\ell}(\hat{q}\cdot\hat{q}_1)\delta_{\ell \ell'} ,
\end{equation}
we can evaluate the $d\Omega_k$ integral to obtain
\begin{equation}
C_{\ell} = \left(\frac{128 \pi^2 m_\nu \Gamma_\nu a^2}{\rho_c}\right) \int \frac{\dbar^3 q dq_1}{8\epsilon_\nu\epsilon_1 \epsilon_2 } q_1^3P_{\ell}(\hat{q}_1.\hat{q}) \Delta f_{\nu\ell}(q) (2\pi)\delta(\epsilon_\nu -\epsilon_1-\epsilon_2).    
\end{equation}
Now, notice that the direction of the neutrino momentum only enters the integrand via the angle between the neutrino momentum $q$ and the daughter momentum $q_1$, defined as $\cos \theta_{1}\equiv \hat{q}\cdot\hat{q}_1$. The energy conserving delta function can be expressed in terms of this angle as
\begin{equation}
 \delta(\epsilon_\nu -\epsilon_1-\epsilon_2)= \frac{\epsilon_2}{q q_1}\delta\left(\cos\theta_{1}-\cos \theta^*_{1}\right),
\end{equation}
where
\begin{equation}
  \cos \theta^*_{ 1}= \frac{2\epsilon_\nu q_1-a^2 m_\nu^2 }{2 q q_1}.
\end{equation}
The energy conservation restricts the daughter momentum to a range of values $(q_1^+,q_1^-)$. The edges of this range occur when the extreme values, $\cos \theta^*_{1}=\pm 1$, are reached. For these values, 
\begin{equation}
   q_{1}^{\pm}=\frac{m_{\nu}^{2} a^{2}}{2\left(\epsilon_{\nu} \pm q\right)}.
\end{equation}
After integrating over the delta function corresponding to energy conservation, this reduces to the simpler form, 
\begin{equation}\label{eq:c_ell2}
    C_{\ell} = \left(\frac{8 \pi m_\nu \Gamma_\nu a^2}{\rho_c}\right) \int \frac{dq}{\epsilon_\nu}  q \Delta f_{\nu(\ell)}\int^{q_1^-}_{q_1^+} dq_1 q_1 P_{\ell}\left(\frac{ 2\epsilon_\nu q_1-a^2 m_\nu^2 }{2 q q_1}\right).
\end{equation}
Eq.~(\ref{eq:c_ell2}) may also be obtained by taking the appropriate limit of the more general expression in Ref.~\cite{Barenboim:2020vrr}. The same Boltzmann hierarchy has been derived in the context of warm matter decaying into dark radiation~\cite{Blinov:2020uvz}.

Performing the integral over $q_1$, we can obtain the following expressions for the first few $C_\ell$'s,
 \begin{eqnarray}\label{eq:C_ell}
&& C_0=\frac{4\pi a^2 \Gamma_{\nu} m_\nu }{\rho_c} \int dq q^2 \Delta f_{\nu(0)},\nonumber\\
 && C_1=\frac{4\pi a^2 \Gamma_{\nu} m_\nu }{\rho_c} \int dq \frac{q^3}{\varepsilon_\nu} \Delta f_{\nu(1)},\nonumber\\
&&    C_2=\frac{4\pi a^2 \Gamma_{\nu} m_\nu }{\rho_c} \int dq q^2 g_2(q,\varepsilon_\nu) \Delta f_{\nu(2)},\nonumber\\
    && C_3=\frac{4\pi a^2 \Gamma_{\nu} m_\nu }{\rho_c} \int dq q^2 g_3(q,\varepsilon_\nu) \Delta f_{\nu(3)}.
 \end{eqnarray}
  Here the functions $g_2(q,\varepsilon_\nu)$ and $g_3(q,\varepsilon_\nu)$ are given by,
 \begin{eqnarray}\label{eq:g_2andg_3}
g_2(q, \varepsilon_{\nu}) &\equiv& \frac{5}{2} - \frac{3}{2} \frac{\varepsilon_{\nu}^2}{q^2} + \frac{3}{4} \frac{(\varepsilon_{\nu}^2-q^2)^2}{\varepsilon_{\nu} q^3} \text{ln} \left( \frac{\varepsilon_{\nu}+q }{\varepsilon_{\nu}-q} \right),\nonumber\\
g_3(q, \varepsilon_{\nu}) &\equiv& \frac{25}{2}\frac{\varepsilon_\nu}{q} -\frac{4q}{\varepsilon_\nu} - \frac{15}{2} \frac{\varepsilon_{\nu}^3}{q^3} + \frac{15}{4} \frac{(\varepsilon_{\nu}^2-q^2)^2}{ q^4} \text{ln} \left( \frac{\varepsilon_{\nu}+q }{\varepsilon_{\nu}-q} \right).
\label{eq:gfunction}
\end{eqnarray}

Given the complicated integrals in Eq.~(\ref{eq:c_ell2}), it is technically challenging to keep track of all the collision terms in the Boltzmann hierarchy. Instead, we choose to keep just the first few $C_\ell$'s for $\ell \leq \ell_{\rm max}$. 
The idea behind this approach is that $C_\ell$ is of  $O((T_{\rm dec}/m_\nu)^\ell)$ around the time of decay. 
Therefore, for non-relativistic decay ($T_{\rm dec}/m_\nu\ll 1$), it is self-consistent to  set $C_{\ell >\ell_{\rm max}}=0$ because those terms only have negligible effect on physical observables. 
To understand the scaling of $C_\ell$, we first note that the integral over $q$ in Eq.~(\ref{eq:C_ell}) receives most of its support from the region around $q\sim T_{\nu0}$ because $\Delta f_{\nu(\ell)}$ inherits features of the Fermi-Dirac distribution from $\bar f_{\rm ini}=1/(e^{q/T_{\nu0}}+1)$. Deep in the non-relativistic region, $q\ll \epsilon_\nu$ and $T_{\nu0} \ll a m_\nu $. In this regime, we can employ a Taylor expansion for the functions $g_2$ and $g_3$ in powers of $q/\epsilon_\nu$ to obtain,
\begin{eqnarray}\label{eq:g_23_approx}
g_2(q,\epsilon_\nu)\approx \frac{4}{5}\frac{q^2}{\epsilon_\nu^2}~~~~,~~~~g_3(q,\epsilon_\nu){\approx} \frac{4}{7}\frac{q^3}{\epsilon_\nu^3} ~~~~~\textrm{for}~(q\ll \epsilon_\nu).
\end{eqnarray}
 Inserting Eq.~(\ref{eq:g_23_approx}) above into Eq.~(\ref{eq:C_ell}), it is straightforward to see that  $C_\ell\propto (T_{\nu0}/a m_\nu)^\ell$.
Moreover, if we assume decay happens deep in the non-relativistic region, we will get $C_\ell\propto (T_{\rm dec}/ m_\nu)^\ell$ when decay happens, where $T_{\rm dec}=T_{\nu0}/a_{\rm dec}$. Therefore, $C_\ell$ is suppressed by powers of $T_{\rm dec}/m_\nu\ll 1$ for higher $\ell$.
To further justify this argument, we show in section~\ref{sec:compare_prescriptions} that setting $\ell_{\rm max}=2$ or $\ell_{\rm max}=3$ makes negligible difference to cosmological observables (see  Fig.~\ref{fig:spectrum_prescription_compare}). Therefore, we only keep $C_{\ell\leq 3}$ and set $C_{\ell>3}=0$ in our numerical study for simplicity.

Physically, the expansion in the small parameter $T_{\rm dec}/m_\nu$ corresponds to perturbing about the ultra-nonrelativistic limit in which the momentum of the mother particle has completely redshifted away, so that it has come to rest in the cosmic frame. Energy and momentum conservation is respected order by order in this expansion. The earlier work \cite{Chacko:2019nej} approximated the Boltzmann hierarchy for daughter radiation (Eq.~\ref{eq:F_r}) by just keeping $C_0$ and setting all the $C_{\ell\geq 1}=0$. It is clear from the above discussion that this is a consistent approximation to zeroth order in an expansion in the small parameter $T_{\rm dec}/m_\nu$. 
The authors in Ref.~\cite{Barenboim:2020vrr} argued that the Boltzmann hierarchy for daughter radiation in Ref.~\cite{Chacko:2019nej} does not reproduce the standard decaying CDM scenario and does not respect momentum conservation. Both criticisms can be addressed by considering the term $C_1$. Since  $C_1$ begins at $O(T_{\rm dec}/m_\nu)$, we see that the Boltzmann hierarchy in Ref.~\cite{Chacko:2019nej} does in fact reproduce the decaying CDM scenario and respects momentum conservation up to $O(T_{\rm dec}/m_\nu)$ corrections, consistent with the approximation.
In this limit, the momenta of the daughter particles arise entirely from the rest mass of the mother. 
In practice, since the contributions of
neutrinos to the density perturbations are small, we will see that the higher order terms do not significantly affect the constraints derived in Ref.~\cite{Chacko:2019nej} with \Planck~2015 data.

\subsection{Signatures of the non-relativistic neutrino decay on the CMB spectra }

\begin{figure}[h]
    \centering
    \includegraphics[width=0.95\columnwidth]{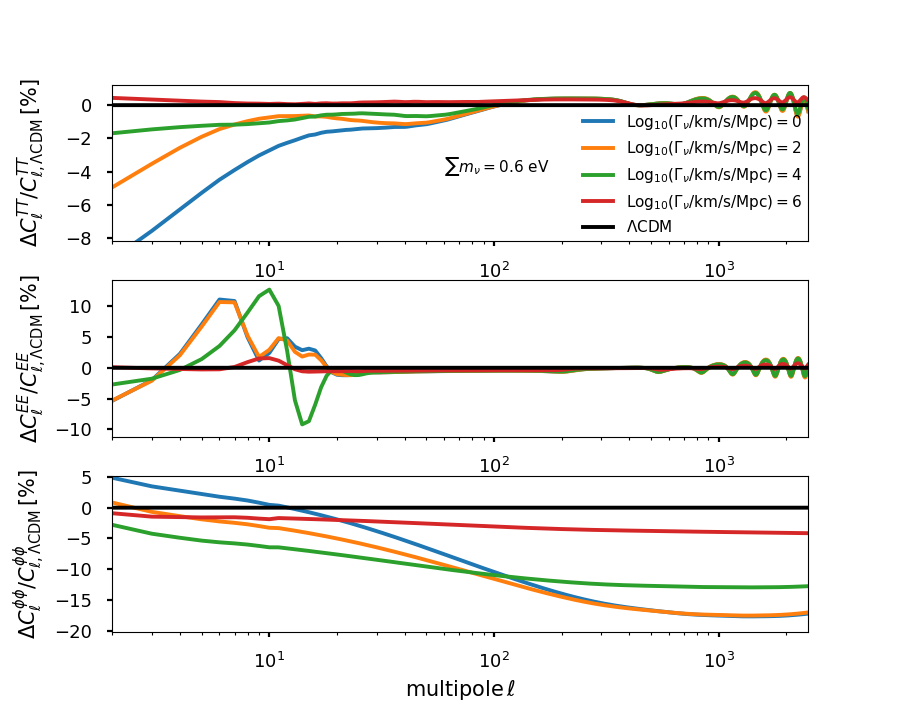}  
    \caption{ Residuals in the the CMB lensed TT (upper), EE (middle) and lensing (lower) spectrum for a fixed value of the neutrino mass and several decay widths. The residuals are taken with respect to the $\Lambda$CDM best-fit parameters from {\it Planck} 2018. The $\Lambda$CDM parameters are kept fixed in all cases.}
    \label{fig:spectrum_gamma_compare}
\end{figure}

To make this work fully self-contained, we briefly summarize the impact of the non-relativistic invisible neutrino decays on the CMB spectra, following the discussion in Ref.~\cite{Chacko:2019nej}.
In Fig. \ref{fig:spectrum_gamma_compare}, we display the residuals in the CMB (lensed) TT, EE and lensing power spectra, for the sum of neutrino masses $\sum m_\nu = 0.6$ eV and several decay widths $\mathrm{Log}_{10}(\Gamma_\nu / \mathrm{km}/\mathrm{s}/\mathrm{Mpc}) = 0,2,4,6$. In all cases, the $\Lambda$CDM parameters are set to their best-fit values from {\it Planck} 2018, that is, $\{100\theta_s =1.04089 $, $\omega_{\rm cdm} = 0.1198$, $\omega_{\rm b} = 0.02233$, $n_s=0.9652$, $\text{ln}(10^{10} A_s)=3.043$, $\tau_{\rm reio}=0.0540\}$. Our reference $\Lambda$CDM model makes use of the same parameters and assumes standard massless neutrinos. 

For the value of the mass considered ($\sum m_\nu = 0.6$ eV) 
and at fixed angular size of the sound horizon $\theta_s$,  neutrino masses primarily impact the lensing spectrum. Indeed, as they reduce power below the free-streaming scale, they produce a significant matter power suppression at small scales, which leads to a $\sim 20 \%$ reduction in the $C_{\ell}^{\phi \phi}$ at large $\ell$  (blue curve in Fig. \ref{fig:spectrum_gamma_compare}). 
Consequently, this power suppression decreases the smoothing  in the high-$\ell$ part of the TT and EE spectra, which can be seen as `wiggles' in the corresponding plots.  

In addition, stable neutrinos  dilute like non-relativistic matter at late times ($\bar{\rho}_{\nu} \sim a^{-3}$), which increases the value of $\Omega_m$. As we impose the closure relation $\Omega_m+\Omega_\Lambda =1$ at late-times, this is compensated for by a decrease in $\Omega_{\Lambda}$ (later beginning of $\Lambda$-domination), and thus a reduction in the Late Integrated Sachs-Wolfe effect (LISW), leaving a signature in the low-$\ell$ TT spectrum. Furthermore, the modified expansion history $H(z)$ changes quantities integrated along $z$, such as $\tau_{\rm reio}$, which affects the multipoles at $\ell \sim 10$ in the EE spectrum.

When a non-negligible $\Gamma_{\nu}$ is considered (orange, green and red curves in Fig. \ref{fig:spectrum_gamma_compare}), one can see that the aforementioned effects typically become less prominent for earlier decays. 
This is particularly true for the high-$\ell$ part of the lensing spectrum (and consequently the smoothing at high-$\ell$ in TT and EE) since decay of neutrinos reduce their impact on structure formation. 
The reduction of the effect in the low-$\ell$ part of the TT and EE spectra is not entirely monotonic, as intermediate values of $\Gamma_{\nu}$ can induce additional time variation in the gravitational potentials (thereby affecting the LISW effect), as well as time variations in $H(z)$ (thereby affecting $\tau_{\rm reio}$). 
As a result, the $\Lambda$CDM limit is reached not only for small values of $\sum m_{\nu}$, but also for high values of $\Gamma_{\nu}$. This will be reflected in the MCMC analysis in section~\ref{sec:MCstudy}, which shows a large positive correlation between both parameters. 
It is precisely this degeneracy which relaxes the neutrino mass bounds.  

\subsection{Consistency of the implementation of Boltzmann equations}
\label{sec:compare_prescriptions}
We begin by comparing the approximation used in Ref.~\cite{Chacko:2019nej} for the background energy density of decaying massive neutrinos to the more accurate results obtained by evaluating the integral in Eq.~(\ref{eq:formal}) numerically. 
In Ref.~\cite{Chacko:2019nej}, the phase space distribution of neutrinos in Eq.~(\ref{eq:formal}) is approximated through the following analytic formula,
\begin{eqnarray}\label{eq:approx_psd}
\bar f_{\nu}(q,\tau)=\bar f_{\rm ini}(q)e^{-\Gamma_{\nu} t/\gamma}.
\end{eqnarray}

\begin{figure}[h]
    \centering
    \includegraphics[width=0.8\columnwidth]{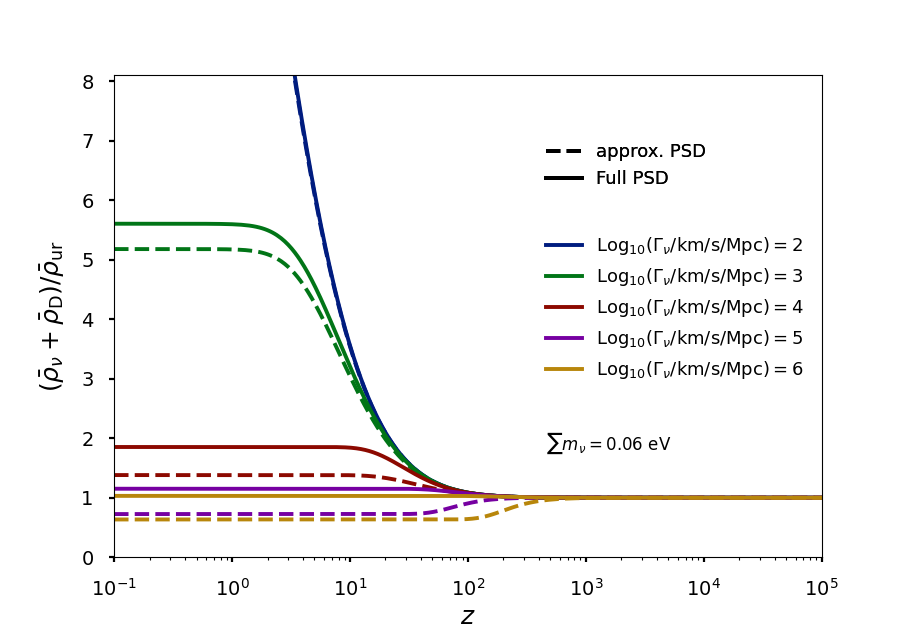} 
    \caption{Redshift evolution of the quantity $(\bar{\rho}_\nu + \bar{\rho}_{\rm D})/\bar{\rho}_{\rm ur}$  (where $\bar{\rho}_{\rm ur}$ denotes the energy density of  stable massless neutrinos), which should be equal to 1 in the limit of relativistic decays. We consider a very small value of the neutrino mass sum, $\sum m_{\nu} = 0.06 \ \rm{eV}$, and several values for the decay width, $\mathrm{Log}_{10}(\Gamma_\nu / \mathrm{km}/\mathrm{s}/\mathrm{Mpc})$. ``approx. PSD'' refers to the approximated phase space distribution in Eq.~(\ref{eq:approx_psd}) while ``Full PSD'' refers to the exact solutions of Eq.~(\ref{eq:formal}).  }
    \label{fig:background_evolution}
\end{figure}

\begin{figure}[h]
    \centering
    \includegraphics[width=0.95\columnwidth]{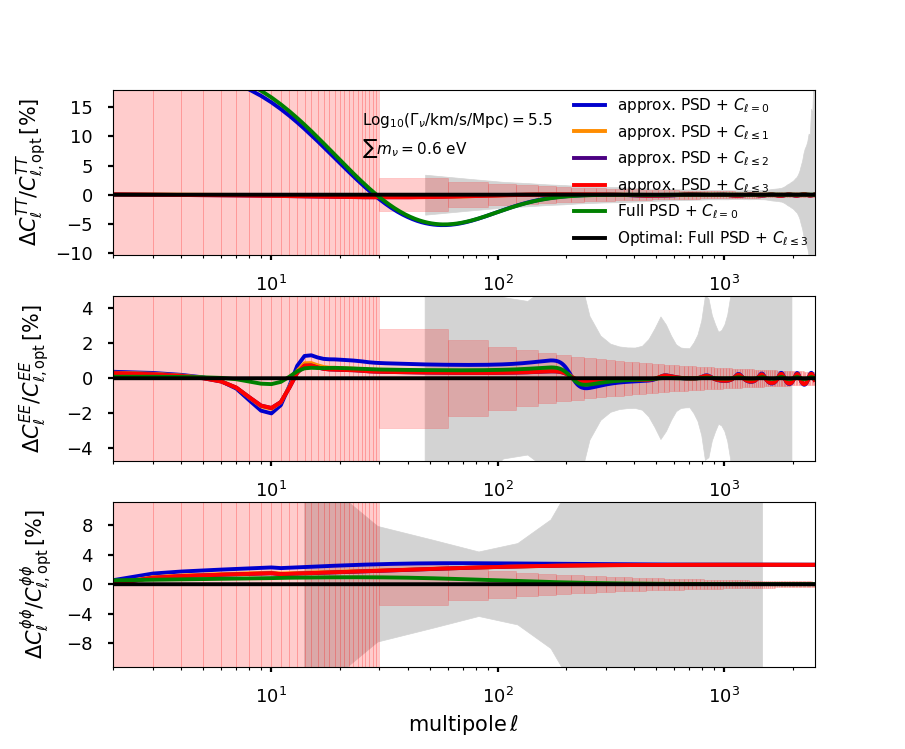}  
    \caption{Fractional change in the CMB TT (upper), EE (middle) and lensing (lower) spectrum, when imposing different prescriptions for the background energy density distribution and Boltzmann hierarchies.  ``approx. PSD'' refers to the approximate phase space distribution in Eq.~(\ref{eq:approx_psd}) while ``Full PSD'' refers to the exact solution of Eq.~(\ref{eq:formal}). ``$C_\ell$'' in the plot means we only keep those collision terms in Eq.~(\ref{eq:F_r}). The chosen values of the neutrino mass ($\sum m_\nu  = 0.6 \ \rm{eV}$) and decay width (${\rm Log}_{10}(\Gamma_{\nu}/[{\rm km/s/Mpc}]) = 5.5$) correspond to the case when neutrinos decay  close to non-relativistic transition ($T_{\rm dec}/m_{\nu} \simeq 0.3$). The gray shaded region indicates {\it Planck} 2018 1-$\sigma$ uncertainties, while the pink boxes indicate the (binned) cosmic variance.}

    \label{fig:spectrum_prescription_compare}
\end{figure}

\begin{figure}[h]
    \centering

    \includegraphics[width=0.95\columnwidth]{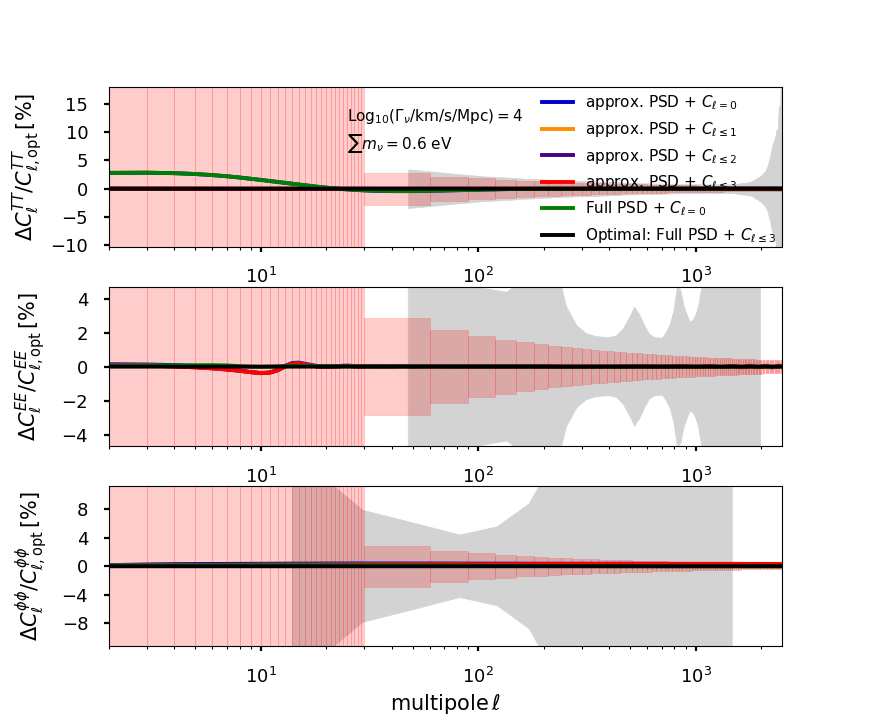}  
    \caption{Same as in Fig. \ref{fig:spectrum_prescription_compare} , but with a smaller decay width (${\rm Log}_{10}(\Gamma_{\nu}/[{\rm km/s/Mpc}]) = 4$), corresponding to a  neutrino decay happening deep in the non-relativistic limit ($T_{\rm dec}/m_{\nu} \simeq 0.03$). }

    \label{fig:spectrum_prescription_compare_2}
\end{figure}

As argued in Ref.~\cite{Chacko:2019nej}, this approximation is valid under the assumption that the decay happens deep in the non-relativistic regime. To see the difference between the approximation and the full result, we plot the ratio $r \equiv (\bar{\rho}_\nu + \bar{\rho}_{\rm D})/\bar{\rho}_{\rm ur}$ in Fig.~\ref{fig:background_evolution}, for several values of the decay width $\Gamma_\nu$ and a fixed value of the total neutrino mass $\sum m_\nu = 0.06 \ \rm{eV}$. Here $\bar{\rho}_{\rm ur}$ denotes the energy density of stable massless neutrinos. If neutrinos decay while relativistic, this ratio always gives $r\simeq 1$. However, if the decay happens when the neutrinos are already non-relativistic ($\bar{\rho}_\nu \sim a^{-3}$ ), then the ratio evolves from $r\simeq 1$ to $r \sim a$, and will eventually reach a plateau once all the neutrinos have decayed. From Fig.~\ref{fig:background_evolution}, we can see that the approximate formula in Eq.~(\ref{eq:approx_psd}) gradually improves as we go to smaller decay widths (that is, going deeper into the regime of non-relativistic decays), as expected. The error in the case of neutrinos decaying right around the time of the non-relativistic transition (${\rm Log}_{10}(\Gamma_{\nu}/[{\rm km/s/Mpc}]) \simeq 4$ for $\sum 0.06$ eV) is around $25 \%$.
Nevertheless, as we argue below, the impact on observables is much smaller given that neutrinos only contribute a small fraction of the total energy density for masses considered in this work.
Not surprisingly, the approximate formula fails in the relativistic regime, leading to $r<1$ at late-times. Therefore future work focusing on this regime should make use of the exact formula.

In Figs.~\ref{fig:spectrum_prescription_compare} and \ref{fig:spectrum_prescription_compare_2}, we show the effects of various approximations in dealing with decaying neutrinos (at the background and perturbation level) on the CMB TT, EE and lensing spectra.
We compare the impact of using either the approximated or the exact PSD of neutrinos discussed above, as well as the impact of only keeping $C_{\ell\leq\ell_{\rm max}}$ in the Boltzmann hierarchy of daughter particles in Eq.~(\ref{eq:F_r}), where we vary $\ell_{\rm max}$ from zero to three. 
We show the residuals of these approximations with respect to the `optimal' case (i.e. including all terms up to $\ell_{\rm max} =3$ and the exact background PSD) for a fixed value of the neutrino mass ($\sum m_{\nu} = 0.6 \ \rm{eV}$) and two different decay widths  (${\rm Log}_{10}(\Gamma_{\nu}/[{\rm km/s/Mpc}]) = 5.5$ in Fig.~\ref{fig:spectrum_prescription_compare} and ${\rm Log}_{10}(\Gamma_{\nu}/[{\rm km/s/Mpc}]) = 4$ in Fig.~\ref{fig:spectrum_prescription_compare_2}). Fig.~\ref{fig:spectrum_prescription_compare} corresponds to decays happening around the time of the non-relativistic transition, $T_{\rm dec}/m_{\nu} \simeq 0.3$, where the effects of the approximations are expected to be largest. Fig.~\ref{fig:spectrum_prescription_compare_2} on the other hand refers to decays happening deep in the non-relativistic regime, $T_{\rm dec}/m_{\nu} \simeq 0.03$.  We also show the {\it Planck} 2018 1-$\sigma$ error bars, as well as the (binned) cosmic variance. 

 For decays close to the non-relativistic transition $T_{\rm dec}/m_{\nu} \simeq 0.3$ shown in Fig.~\ref{fig:spectrum_prescription_compare}, we find that the biggest improvement in the CMB TT spectrum occurs when including $C_{\ell \leq 1}$ (i.e., the contribution from the decaying neutrino bulk velocity) in the Boltzmann hierarchy of daughter radiation, which impacts the integrated Sachs-Wolfe (ISW) effect at multipoles $\ell \lesssim 100$. 
On the other hand, the approximate background distribution of neutrinos does not have a significant effect.
For the CMB EE spectrum shown in the same figure, which is not sourced by the ISW effect, the impact of the approximate background distribution of neutrinos is comparable to the effect of the approximate perturbed hierarchy. Nevertheless, one can see that for $\ell_{\rm max}\geq 2$, additional contributions to the daughter hierarchy have negligible impacts, which justifies our choice of cutting the  collision term $C_\ell$ contribution at $\ell_{\rm max}=3$.
Finally for the CMB lensing spectrum, the effects due to the approximate treatment of the background PSD dominate over the ones due to including higher order terms in the Boltzmann hierarchy of the dark radiation. 
This is expected given that the matter power spectrum suppression scales approximately with $\bar{\rho}_\nu/\bar{\rho}_{m}$ \cite{Hu:1997mj,Lesgourgues:2018ncw} where $\bar{\rho}_m$ is the total matter density, while neutrino perturbations are very small well below the free-streaming scale, so that their detailed dynamics is not as important as on larger scales.  

The impact of the various approximations in the case of decays deep in the non-relativistic regime $T_{\rm dec}/m_{\nu} \simeq 0.03$, displayed in Fig.~\ref{fig:spectrum_prescription_compare_2}, is much less visible. In that case, one can therefore safely neglect $C_{\ell>0}$ and consider the approximate PSD, as done in Ref.~\cite{Chacko:2019nej}.

\section{Updated Monte Carlo analysis of the decaying neutrino scenario}\label{sec:MCstudy}

\subsection{Details of the analysis}
In this section we perform a numerical scan over the parameter space to obtain updated limits on the neutrino mass and lifetime.
We perform comprehensive MCMC analyses with the \texttt{MontePython-v3}\footnote{ https://github.com/brinckmann/montepython\_public} \cite{Audren:2012wb,Brinckmann:2018cvx} code interfaced with our modified version of \texttt{CLASS}.
We fit the decaying neutrino model to a combination of the following data-sets:

\begin{itemize}
   \item The {\it Planck} 2018 high-$\ell$ TT, TE, EE + low-$\ell$ data TT, EE +  lensing data~\cite{Aghanim:2018eyx}. We will also compare these results with the use of {\it Planck} 2015 data to disentangle the effects of our improved formalism and that of the new data.
   
    \item  The BAO measurements from 6dFGS at $z=0.106$~\cite{Beutler:2011hx}, SDSS DR7 at $z=0.15$~\cite{Ross:2014qpa}, BOSS DR12 at $z=0.38, 0.51$ and $0.61$~\cite{Alam:2016hwk}, and the joint constraints from eBOSS DR14 Ly-$\alpha$ auto-correlation at $z=2.34$~\cite{Agathe:2019vsu} and cross-correlation at $z=2.35$~\cite{Blomqvist:2019rah}.
    
    \item The measurements of the growth function $f\sigma_8(z)$ (FS) from the CMASS and LOWZ galaxy samples of BOSS DR12 at $z = 0.38$, $0.51$, and $0.61$~\cite{Alam:2016hwk}.
    
    \item The Pantheon SNIa catalogue, spanning redshifts $0.01 < z < 2.3$~\cite{Scolnic:2017caz}.
\end{itemize}

We adopt wide flat priors on the following six $\Lambda$CDM parameters: $\{\omega_b,\omega_{\rm cdm}, H_0,n_s,A_s,\tau_{\rm reio}\}$. 
We assume three degenerate neutrinos decaying into massless radiation and consider flat priors on $\sum m_{\nu}/{\rm eV}$ and ${\rm Log}_{10}(\Gamma_{\nu}/[{\rm km/s/Mpc}])$. 
To accelerate convergence, we split the parameter space between  ${\rm Log}_{10}(\Gamma_{\nu}/[{\rm km/s/Mpc}])\in[0.1,2.5]$ and  ${\rm Log}_{10}(\Gamma_{\nu}/[{\rm km/s/Mpc}])\in[2.5,6.5]$. 
In both cases we take wide priors on $\sum m_\nu\in[0.06,1.5]$ eV.
We assume our MCMC chains to be converged when the Gelman-Rubin criterion $R-1 < 0.05$ \cite{Gelman:1992zz}.  
In our baseline analysis, we do not apply any specific cut to the parameter space, even if neutrinos decay in the relativistic regime (this occurs for low $\sum m_{\nu}$ and high $\Gamma_{\nu}$). 
In appendix \ref{app:mcmc}, we investigate the impact of 
 imposing a prior that excludes the parameter space corresponding to relativistic decay from our analysis and show that the limit at 95\% on $\sum m_{\nu}$ agrees within a few percent.

\subsection{Main results: updated limit on the neutrino mass and lifetime}

\begin{figure*}
    \centering
     \includegraphics[width=1\columnwidth]{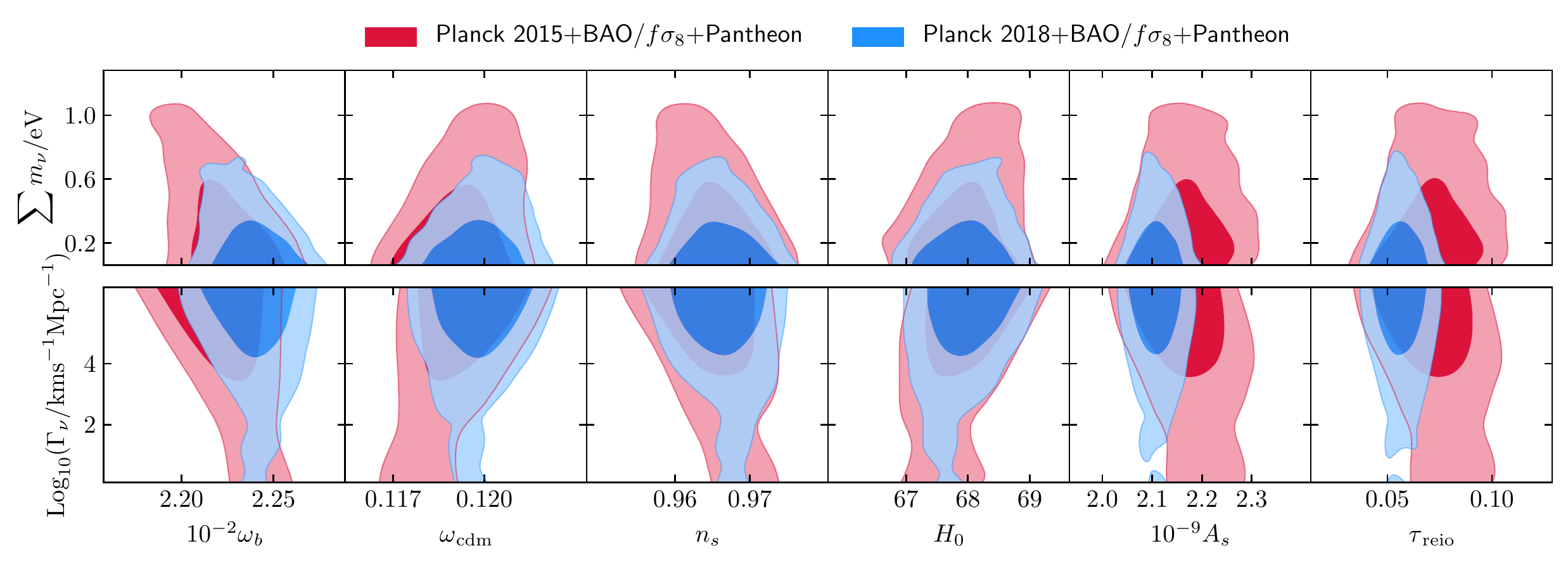}
          \includegraphics[width=0.4\columnwidth]{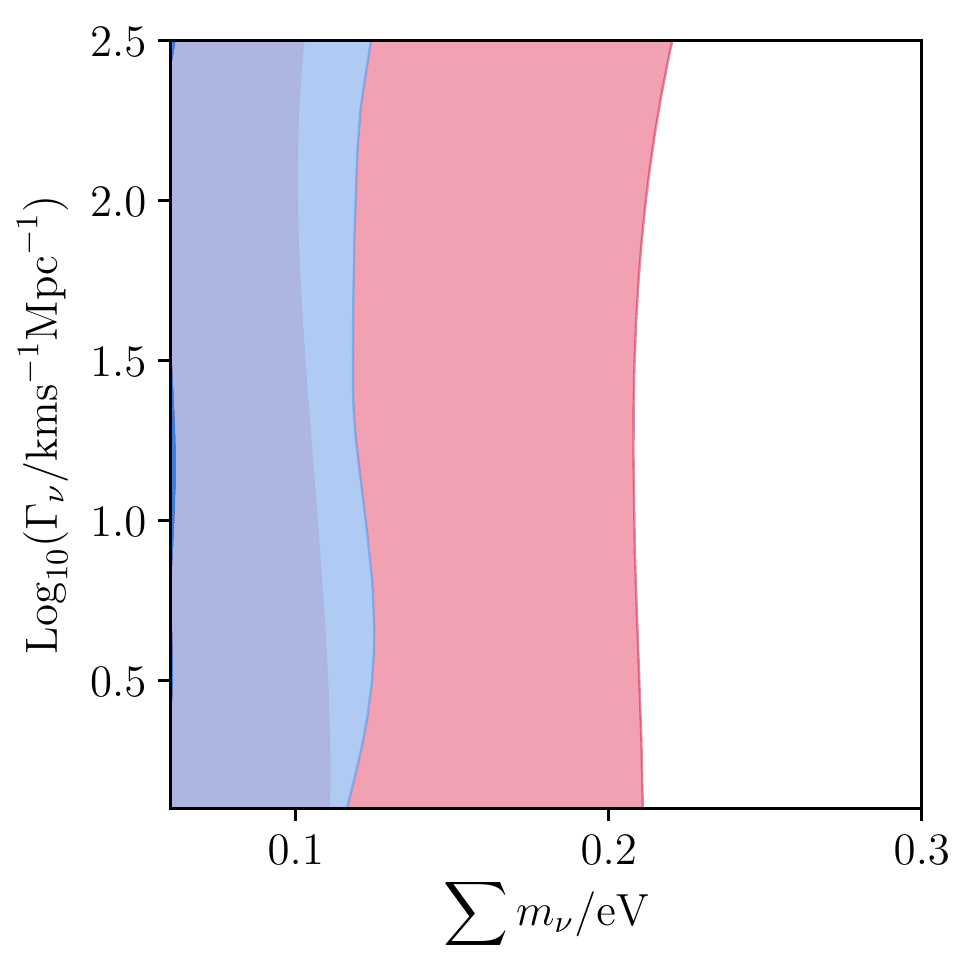}    \includegraphics[width=0.4\columnwidth]{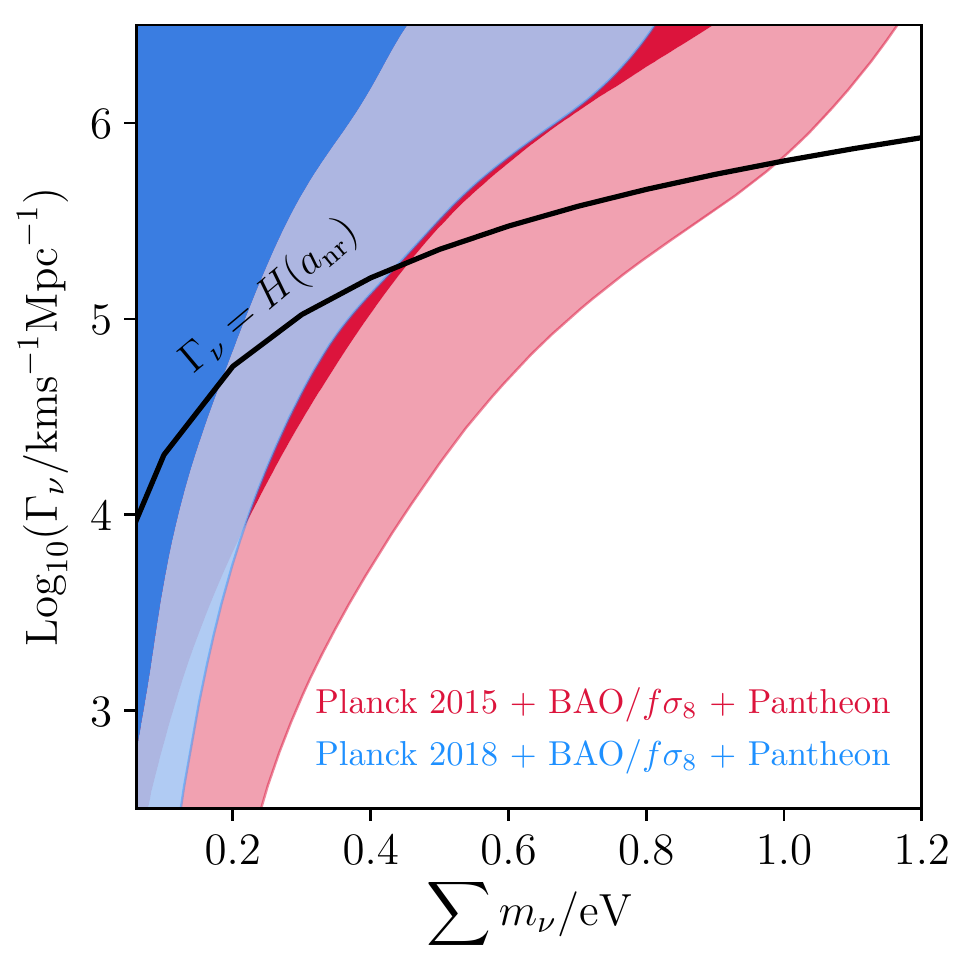}     
    \caption{2D posterior distribution of the decaying neutrino model reconstructed from the analysis of BAO + FS + Pantheon together with either \Planck~2015 or {\it Planck} 2018 data. 
    In the top panel, we show the correlation with other cosmological parameters. } 
    \label{fig:MCMC_results}
\end{figure*}

The results of our analyses are presented in Figs.~\ref{fig:MCMC_results}. 
For very late decays, ${\rm Log}_{10}(\Gamma_{\nu}/[{\rm km/s/Mpc}])\lesssim 2.5$, no relaxation of the constraints on $\sum m_{\nu}/$eV is visible, in agreement with what was found in Ref.~\cite{Chacko:2019nej}. 
The impact of the new {\it Planck} data is visible as a significantly improved bound on the sum of neutrino mass, namely we find $\sum m_{\nu}< 0.127$ eV (95\%C.L.), an improvement of about $\sim 35\%$ over 2015 data, in good agreement with Ref.~\cite{Aghanim:2018eyx}. 
For ${\rm Log}_{10}(\Gamma_{\nu}/[{\rm km/s/Mpc}])\gtrsim 2.5$, one can see that the bound relaxes as expected, although not as much with \Planck~2018 data as for \Planck~2015 data. 

 Taking the intersect of the non-relativistic decay line as our 2$\sigma$ limit, we find that \Planck~2018 allows neutrinos with masses up to $\sum m_{\nu} = 0.42$~eV.   In appendix~\ref{app:mcmc}, we present an alternative analysis that directly imposes the non-relativistic decay criterion as a prior while performing the scan. Marginalizing over all parameters we find the same result, $\sum m_{\nu} < 0.42$ eV (95\% C.L.). The excellent agreement between these different analyses leads us to conclude with confidence that,  within the regime of non-relativistic decay, values of $\sum m_{\nu}$ as large as 0.4 eV are still allowed by the data. This bound is significantly stronger than the limit from \Planck~2015 data, for which $\sum m_{\nu}\sim 0.9$ eV was still allowed in the non-relativistic decay scenario. 

Our result also has implications for laboratory searches. For $\sum m_\nu=0.6$ eV, the smallest mass scale that the KATRIN experiment is designed to probe,  \Planck~2018 data requires decay rate $\Gamma_\nu \gtrsim 10^{5.5}$ km/s/Mpc, a constraint roughly one order of magnitude stronger than from \Planck~2015 data. 
However, this value of the decay rate is now slightly beyond the regime of validity of our work\footnote{For $\sum m_\nu=0.6$ eV and assuming degenerate neutrino masses, the non-relativistic condition requires $\Gamma_\nu < 10^{5.3}$ km/s/Mpc.}, indicating that, in the event of a neutrino mass discovery at KATRIN, a more involved analysis including inverse-decays would be necessary to confirm that the decay scenario can reconcile laboratory and cosmological measurements.

\subsection{Comparison with former results and the impact of {\it Planck} 2018 data}

\begin{figure*}
    \centering
        \includegraphics[width=0.5\columnwidth]{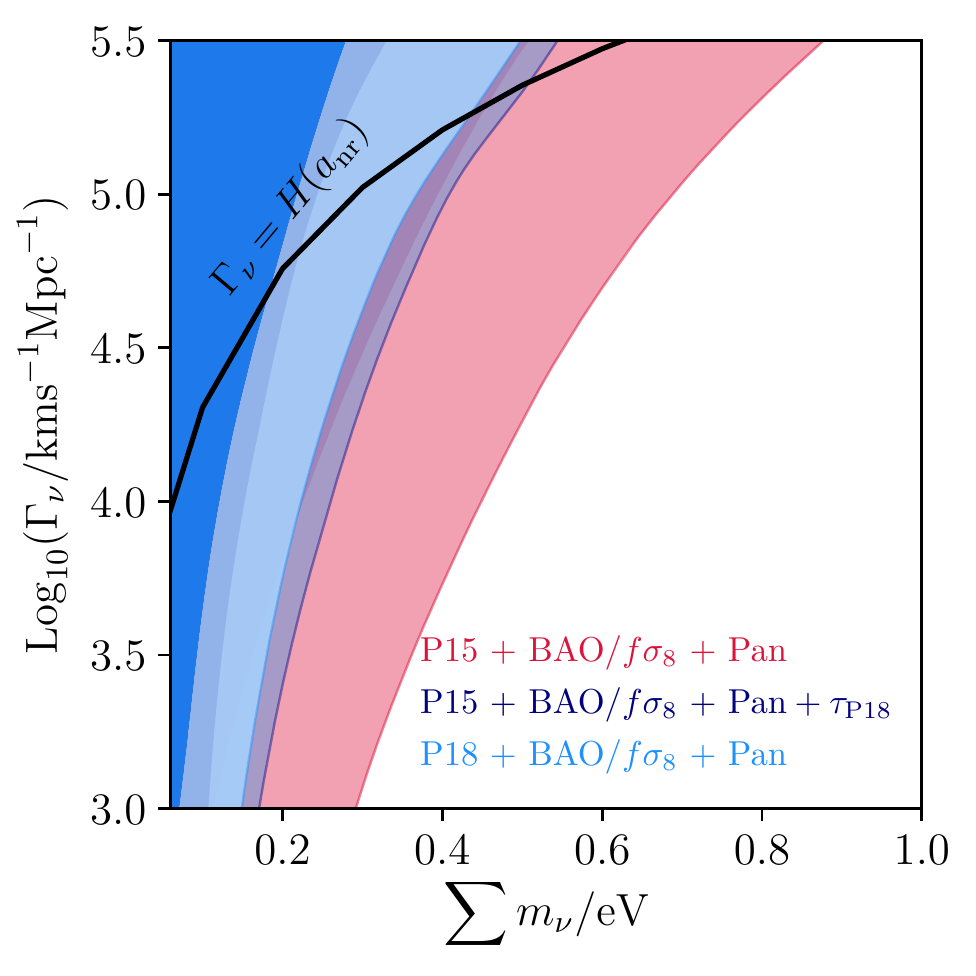}
    \caption{Posterior distribution of $\sum m_{\nu}$ and ${\rm Log}_{10}(\Gamma_{\nu}/[{\rm km/s/Mpc}])$ with {\it Planck} 2018 and {\it Planck} 2015 + a primer on $\tau_{\rm reio}$ from {\it Planck} 2018 . The agreement of the posteriors shows the dominant constraining power on $\sum m_{\nu}$ and ${\rm Log}_{10}(\Gamma_{\nu}/[{\rm km/s/Mpc}])$ comes from a precise measurement of $\tau_{\rm reio}$ from {\it Planck} 2018. 
    }
    \label{fig:tau_primer}
\end{figure*}

Comparing with the constraints presented in Ref.~\cite{Chacko:2019nej} for \Planck~2015, we find that, while the impact of our improved treatment is clearly visible in the CMB power spectra (and will be relevant for future experiments), it has only a marginal impact on the constraints, and our bounds are in very good agreement with those derived in Ref.~\cite{Chacko:2019nej}, which only included the leading order term in the daughter radiation hierarchy\footnote{Let us note that the implementation of the BAO/$f\sigma_8$ DR12 likelihood used in Ref.~\cite{Chacko:2019nej} within the MontePython code had an issue that led to constraints on $\sum m_\nu$ that were somewhat milder than the true bounds. MontePython has since then been corrected, leading to an improvement on the constraints on the stable/long-lived ($\Gamma_\nu<10^3$ km/s/Mpc) case by about $20\%$. However, we have verified that this bug had no impact in the short-lived case ($\Gamma_\nu>10^{2.5}$ km/s/Mpc).}. 
The bulk of the improvement is due to the newest \Planck~2018 data and can be understood as follows.
As shown in Fig.~\ref{fig:spectrum_gamma_compare}, for the masses we consider, the main effect  is an  almost scale independent suppression of CMB lensing spectrum. 
This suppression can be compensated for by increasing the primordial amplitude $A_s$ or by adjusting the matter density $\omega_{\rm cdm}$ (see Ref.~\cite{Archidiacono:2016lnv} for a discussion of the correlation between $\{\sum m_\nu,A_s,\tau_{\rm reio},\omega_{\rm cdm}\}$). 
Due to the well-known degeneracy between $A_s$ and $e^{-2\tau_{\rm reio}}$, {\it Planck} 2015 data, which was limited in polarization, were unable to place a tight constraint on $A_s$, and thus the constraining power on the sum of neutrino mass and lifetime was limited.
 The precise  measurements  of  low-$\ell$ polarization  from  {\it Planck}  2018 leads to constraints on $\tau_{\rm reio}$ that are tighter by a factor of two than those from {\it Planck} 2015. 
 As a result, parameters degenerate with $\tau_{\rm reio}$ such as $A_s$ are now much better constrained. 
 Consequently, the constraints on the sum of neutrino mass and lifetime have significantly improved with {\it Planck} 2018 data. 
 To confirm this simple argument, we perform another MCMC run with {\it Planck} 2015 data and a tight gaussian prior on $\tau_{\rm reio} = 0.0540 \pm 0.0074$, chosen to match the optical depth to reionization reconstructed from {\it Planck} 2018. 
 Given that the constraints on $\sum m_\nu$ are independent of $\Gamma_\nu$ below $\Gamma_\nu\lesssim10^3$, and the scaling above $\Gamma_\nu\lesssim10^{5.5}$ is monotonic, we focus on the parameter space ${\rm Log}_{10}(\Gamma_{\nu}/[{\rm km/s/Mpc}])\in[3,5.5]$ to accelerate convergence.
 Our results are presented in Fig.~\ref{fig:tau_primer}, where one can see that this simple prescription leads to constraints that are very similar to those from the  full {\it Planck} 2018 data. 
 We attribute the remaining differences to the additional constraining power of \Planck~2018 data on the parameters $\omega_{\rm cdm}$ and $\omega_b$, which are mildly correlated with $\sum m_\nu$ (see Fig.~\ref{fig:MCMC_results}, top panel). 
Note that our constraints are a factor of two weaker than those advocated in Ref.~\cite{Lorenz:2021alz}, which performed a `model-independent' reconstruction of the neutrino mass as a function of redshift, but neglects the decay products.
 As we show here, including details about the daughter radiation is necessary to accurately compute the effect of neutrino decays even in the non-relativistic regime.
Finally, as discussed in Refs.~\cite{Archidiacono:2016lnv,Chacko:2020hmh}, a combination of CMB data with future tomographic measurements of the power spectrum by DESI \cite{Font-Ribera:2013rwa} or Euclid \cite{Amendola:2016saw}, and an improved determination of the optical depth to reionization by 21-cm observations with SKA \cite{Liu:2015txa,Maartens:2015mra}, could greatly increase the sensitivity of cosmological probes to neutrino masses and lifetimes.

\section{Conclusions}\label{sec:conclusions}
Cosmological observations are known to set the strongest constraints on the sum of neutrino masses.  
Yet, the existing mass bound from CMB and LSS measurements, which assumes that neutrinos are stable, is significantly weakened if neutrinos decay.
In this work, we provide up-to-date limits  on the lifetime of massive neutrinos that decay into dark radiation after becoming non-relativistic, from a combination of CMB, BAO, growth factor measurements, and Pantheon SN1a data.

Compared to the earlier analysis~\cite{Chacko:2019nej}, we have incorporated higher-order corrections up to ${O}((T_{\rm dec}/m_{\nu})^3)$ when solving the dark radiation perturbations, and also performed the full calculation of the background energy density of the decaying
neutrino using Eq.~(\ref{eq:formal}). 
The more precise treatment of the Boltzmann equations and the background energy evolution in our MCMC study improves the coverage of the case when the neutrinos decay early so that their average momenta are close to their masses.
As shown in Fig.~\ref{fig:spectrum_prescription_compare_2}, if neutrinos decay when having $T_\nu\ll m_\nu/3$, the inclusion of higher moment perturbations $C_{\ell\geq2}$ gives a negligible change to the power spectra as compared to the experimental uncertainties. 
However, the complete calculation of the neutrino energy does improve the prediction for the power spectrum significantly from the approximate result using Eq.~(\ref{eq:approx_psd}) when the decays happen semi-relativistically. 
Nevertheless, we have found that constraints from \Planck~2015, given their limited precision, are unaffected by these considerations. However, we anticipate that these effects will be relevant for future experiments (as well as an essential contribution in the relativistic case, to be considered in the future).

In fact, we have shown that the bulk of the improvement in the constraining power compared to Ref.~\cite{Chacko:2019nej} comes from the use of \Planck~2018 data. 
Indeed, we have demonstrated that the improved $\tau_{\rm reio}$ measurement from the low-$\ell$ polarization data helps breaking the degeneracy in the CMB power spectrum amplitude and strengthens the bound on the neutrino mass and lifetime.  
As a result, we have found that neutrinos with $\sum m_{\nu}>0.42$~eV ($2\sigma$) cannot be made consistent with cosmological data if they decay while non-relativistic, a significant improvement from \Planck~2015 data for which masses as high as $\sum m_\nu \sim 0.9$ eV were consistent with the non-relativistic decay scenario \cite{Chacko:2019nej}. 

We have argued that one notable application of this result is that, if the KATRIN experiment sees an electron neutrino with $m_{\nu}\approx 0.2$~eV (the advocated sensitivity), our result would constrain $\Gamma_\nu \gtrsim 10^{5.5}$ km/s/Mpc, i.e. the neutrinos would need to decay between $z\approx2\times 10^2- 4\times 10^3$, while they are still relativistic, so that our bounds and the bounds studied in Ref.~\cite{Barenboim:2020vrr} would not apply.
In case of a neutrino mass discovery at KATRIN, a more involved analysis including inverse-decays would be necessary to firmly confirm that the decay scenario can reconcile laboratory and cosmological measurements. 
Additionally, our results show that the tentative exclusion of the inverted mass ordering \cite{Vagnozzi:2017ovm,Simpson:2017qvj,DiValentino:2021hoh,Palanque-Delabrouille:2019iyz}, based solely on the fact that the inverted ordering predicts $\sum m_\nu > 0.1$ eV, is highly dependent on the hypothesis that neutrinos are stable on cosmological time-scales. 
Non-relativistic decays can still easily reconcile the inverted ordering with cosmological data.

Finally, let us mention that even though current exclusion bounds in Fig.~\ref{fig:MCMC_results} do not set independent constraints on the neutrino mass and lifetime, next generation measurements of the matter power spectrum at different redshifts can help break that degeneracy~\cite{Chacko:2020hmh}.  
It will be interesting to revisit the forecast on the sensitivity of future cosmological data to the sum of neutrino masses and their lifetime in light of our improved formalism.

\section*{Acknowledgements}

PD is supported in part by NSF grant PHY-1915093. ZC is supported in part by the National Science Foundation under Grant Number PHY-1914731. ZC is also supported in part by the US-Israeli BSF grant 2018236.  YT is supported by the NSF grant PHY-2014165 and PHY-2112540. Fermilab is operated by Fermi Research Alliance, LLC
under contract number DE-AC02-07CH11359 with the United States Department of Energy. This work has been partly supported by the CNRS-IN2P3 grant Dark21. The authors acknowledge the use of computational resources from the Dark Energy computing Center funded by the Excellence Initiative of Aix-Marseille University - A*MIDEX, a French "Investissements d'Avenir" programme (AMX-19-IET-008 - IPhU). PD and YT thank the Aspen Center for Physics, which is supported by National Science Foundation grant PHY- 1607611, where part of this work was performed. PD was also partially supported by a grant from the Simons Foundation during the stay at the Aspen Center for Physics.

\appendix

\section{Excluding the relativistic decay regime from the MCMC analysis}
\label{app:mcmc}
In our baseline analysis, we have extrapolated our scans to the (mildly-)relativistic decay regime, despite the fact that the equations do not include inverse decays. We have then interpreted the bound on the sum of neutrino masses when considering non-relativistic decays as the intersect between the non-relativistic decay condition $\Gamma_\nu > H(T_{\nu}=m_{\nu}/3)$  and the $2\sigma$ limit derived from our analysis. 

\begin{figure}
    \centering
    \includegraphics[width=0.5\columnwidth]{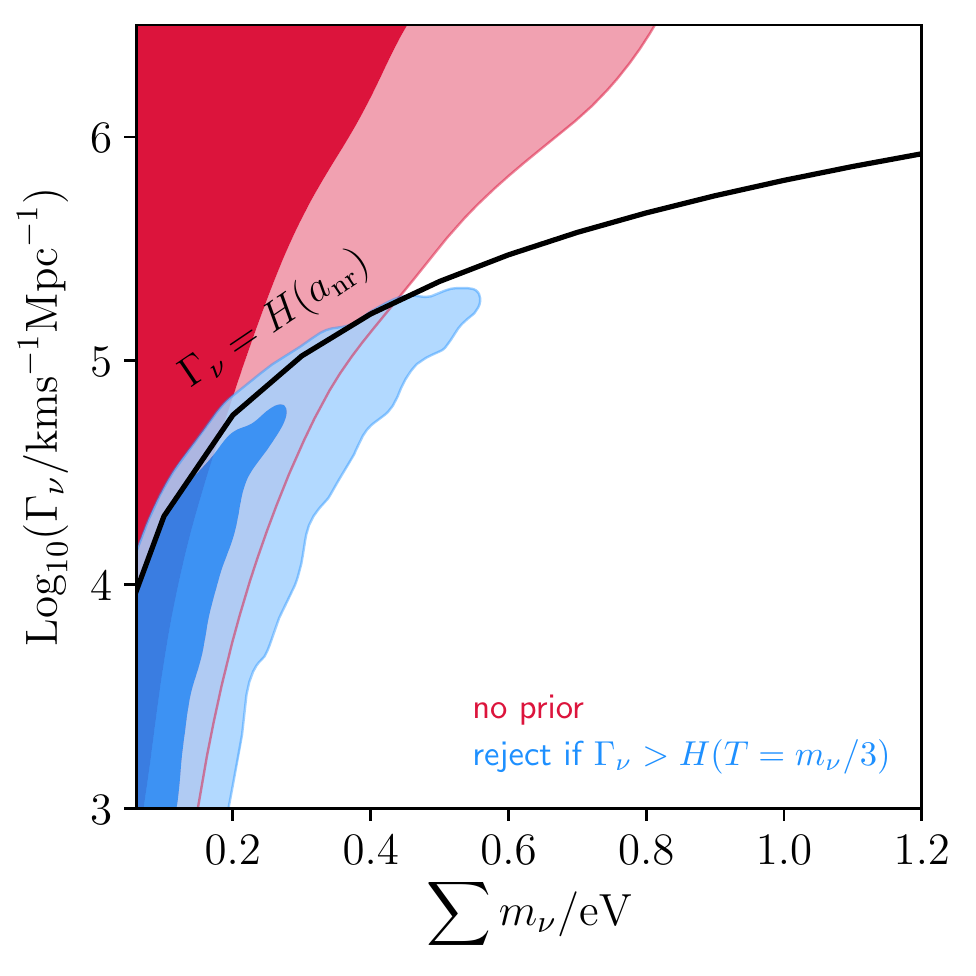}
    \caption{Posterior distribution of $\sum m_{\nu}$ and ${\rm Log}_{10}(\Gamma_{\nu}/[{\rm km/s/Mpc}])$ when confronted to {\it Planck} 2018 + BAO + FS + Pantheon for two different choices of priors on $\Gamma_{\nu}$ (see legend).
    }
    \label{fig:mcmc_priors}
\end{figure}

In this appendix, we investigate how excluding the relativistic decay regime of parameter space from the scan can affect the bounds on $\sum m_{\nu}/$eV and ${\rm Log}_{10}(\Gamma_{\nu}/[{\rm km/s/Mpc}])$. 
As we are interested in (semi-)relativistic decays, we focus on the parameter space ${\rm Log}_{10}(\Gamma_{\nu}/[{\rm km/s/Mpc}])\in[3,6.5]$.
Our results are presented in Fig.~\ref{fig:mcmc_priors}.
In the 2D plane $\{{\rm Log}_{10}\Gamma_{\nu},\sum m_{\nu}\}$ and below the non-relativistic line $\Gamma_\nu = H(T_{\nu}=m_{\nu}/3)$, we find that imposing the condition directly within the MCMC prior relaxes the bound by $\sim 10-20\%$. Nevertheless, after marginalizing over ${\rm Log}_{10}(\Gamma_{\nu})$),  we find that the `naive' bound coming from the intersect between the non-relativistic line ( $\Gamma_\nu > H(T_{\nu}=m_{\nu}/3)$) and the 2$\sigma$ limit without priors is in excellent agreement with that coming from imposing this condition as a prior in the analysis, both yielding $\sum m_{\nu}< 0.42$ eV.

\addcontentsline{toc}{section}{Bibliography} 
\bibliography{references_fluid_neutrinos} 

\providecommand{\noopsort}[1]{}\providecommand{\singleletter}[1]{#1}%
\begin{thebibliography}{10}

\bibitem{Bond:1980ha}
J.~R. Bond, G.~Efstathiou, and J.~Silk, ``{Massive Neutrinos and the Large
  Scale Structure of the Universe},'' {\em Phys. Rev. Lett.}, vol.~45,
  pp.~1980--1984, 1980.
\newblock [,61(1980)].

\bibitem{Hu:1997mj}
W.~Hu, D.~J. Eisenstein, and M.~Tegmark, ``{Weighing neutrinos with galaxy
  surveys},'' {\em Phys. Rev. Lett.}, vol.~80, pp.~5255--5258, 1998.

\bibitem{Wong:2011ip}
Y.~Y.~Y. Wong, ``{Neutrino mass in cosmology: status and prospects},'' {\em
  Ann. Rev. Nucl. Part. Sci.}, vol.~61, pp.~69--98, 2011.

\bibitem{Lesgourgues:2018ncw}
J.~Lesgourgues, G.~Mangano, G.~Miele, and S.~Pastor, {\em {Neutrino
  Cosmology}}.
\newblock Cambridge University Press, 2 2013.

\bibitem{Tanabashi:2636832}
M.~Tanabashi {\em et~al.}, ``{Review of Particle Physics},'' {\em Phys. Rev.
  D}, vol.~98, no.~3, p.~030001. 1898 p, 2018.

\bibitem{Lattanzi:2017ubx}
M.~Lattanzi and M.~Gerbino, ``{Status of neutrino properties and future
  prospects - Cosmological and astrophysical constraints},'' {\em Front.in
  Phys.}, vol.~5, p.~70, 2018.

\bibitem{Aghanim:2018eyx}
N.~Aghanim {\em et~al.}, ``{Planck 2018 results. VI. Cosmological
  parameters},'' {\em Astron. Astrophys.}, vol.~641, p.~A6, 2020.

\bibitem{Serpico:2007pt}
P.~D. Serpico, ``{Cosmological neutrino mass detection: The best probe of
  neutrino lifetime},'' {\em Phys. Rev. Lett.}, vol.~98, p.~171301, 2007.

\bibitem{Serpico:2008zza}
P.~D. Serpico, ``{Neutrinos and cosmology: a lifetime relationship},'' {\em J.
  Phys. Conf. Ser.}, vol.~173, p.~012018, 2009.

\bibitem{Beacom:2004yd}
J.~F. Beacom, N.~F. Bell, and S.~Dodelson, ``{Neutrinoless universe},'' {\em
  Phys. Rev. Lett.}, vol.~93, p.~121302, 2004.

\bibitem{Farzan:2015pca}
Y.~Farzan and S.~Hannestad, ``{Neutrinos secretly converting to lighter
  particles to please both KATRIN and the cosmos},'' {\em JCAP}, vol.~1602,
  no.~02, p.~058, 2016.

\bibitem{Aalberts:2018obr}
J.~L. Aalberts {\em et~al.}, ``{Precision constraints on radiative neutrino
  decay with CMB spectral distortion},'' {\em Phys. Rev.}, vol.~D98, p.~023001,
  2018.

\bibitem{Barenboim:2020vrr}
G.~Barenboim, J.~Z. Chen, S.~Hannestad, I.~M. Oldengott, T.~Tram, and Y.~Y.~Y.
  Wong, ``{Invisible neutrino decay in precision cosmology},'' {\em JCAP},
  vol.~03, p.~087, 2021.

\bibitem{Hannestad:2005ex}
S.~Hannestad and G.~Raffelt, ``{Constraining invisible neutrino decays with the
  cosmic microwave background},'' {\em Phys. Rev. D}, vol.~72, p.~103514, 2005.

\bibitem{Basboll:2008fx}
A.~Basboll, O.~E. Bjaelde, S.~Hannestad, and G.~G. Raffelt, ``{Are cosmological
  neutrinos free-streaming?},'' {\em Phys. Rev. D}, vol.~79, p.~043512, 2009.

\bibitem{Archidiacono:2013dua}
M.~Archidiacono and S.~Hannestad, ``{Updated constraints on non-standard
  neutrino interactions from Planck},'' {\em JCAP}, vol.~07, p.~046, 2014.

\bibitem{Escudero:2019gfk}
M.~Escudero and M.~Fairbairn, ``{Cosmological Constraints on Invisible Neutrino
  Decays Revisited},'' {\em Phys. Rev. D}, vol.~100, no.~10, p.~103531, 2019.

\bibitem{Chacko:2019nej}
Z.~Chacko, A.~Dev, P.~Du, V.~Poulin, and Y.~Tsai, ``{Cosmological Limits on the
  Neutrino Mass and Lifetime},'' {\em JHEP}, vol.~04, p.~020, 2020.

\bibitem{Chacko:2020hmh}
Z.~Chacko, A.~Dev, P.~Du, V.~Poulin, and Y.~Tsai, ``{Determining the Neutrino
  Lifetime from Cosmology},'' {\em Phys. Rev. D}, vol.~103, no.~4, p.~043519,
  2021.

\bibitem{Frieman:1987as}
J.~A. Frieman, H.~E. Haber, and K.~Freese, ``{Neutrino Mixing, Decays and
  Supernova Sn1987a},'' {\em Phys. Lett.}, vol.~B200, pp.~115--121, 1988.

\bibitem{Joshipura:2002fb}
A.~S. Joshipura, E.~Masso, and S.~Mohanty, ``{Constraints on decay plus
  oscillation solutions of the solar neutrino problem},'' {\em Phys. Rev.},
  vol.~D66, p.~113008, 2002.

\bibitem{Beacom:2002cb}
J.~F. Beacom and N.~F. Bell, ``{Do solar neutrinos decay?},'' {\em Phys. Rev.},
  vol.~D65, p.~113009, 2002.

\bibitem{Bandyopadhyay:2002qg}
A.~Bandyopadhyay, S.~Choubey, and S.~Goswami, ``{Neutrino decay confronts the
  SNO data},'' {\em Phys. Lett.}, vol.~B555, pp.~33--42, 2003.

\bibitem{Berryman:2014qha}
J.~M. Berryman, A.~de~Gouvea, and D.~Hernandez, ``{Solar Neutrinos and the
  Decaying Neutrino Hypothesis},'' {\em Phys. Rev. D}, vol.~92, no.~7,
  p.~073003, 2015.

\bibitem{Baerwald:2012kc}
P.~Baerwald, M.~Bustamante, and W.~Winter, ``{Neutrino Decays over Cosmological
  Distances and the Implications for Neutrino Telescopes},'' {\em JCAP},
  vol.~10, p.~020, 2012.

\bibitem{Pagliaroli:2015rca}
G.~Pagliaroli, A.~Palladino, F.~L. Villante, and F.~Vissani, ``{Testing
  nonradiative neutrino decay scenarios with IceCube data},'' {\em Phys. Rev.
  D}, vol.~92, no.~11, p.~113008, 2015.

\bibitem{Bustamante:2016ciw}
M.~Bustamante, J.~F. Beacom, and K.~Murase, ``{Testing decay of astrophysical
  neutrinos with incomplete information},'' {\em Phys. Rev. D}, vol.~95, no.~6,
  p.~063013, 2017.

\bibitem{Denton:2018aml}
P.~B. Denton and I.~Tamborra, ``{Invisible Neutrino Decay Could Resolve
  IceCube\textquoteright{}s Track and Cascade Tension},'' {\em Phys. Rev.
  Lett.}, vol.~121, no.~12, p.~121802, 2018.

\bibitem{Abdullahi:2020rge}
A.~Abdullahi and P.~B. Denton, ``{Visible Decay of Astrophysical Neutrinos at
  IceCube},'' {\em Phys. Rev. D}, vol.~102, no.~2, p.~023018, 2020.

\bibitem{Bustamante:2020niz}
M.~Bustamante, ``{New limits on neutrino decay from the Glashow resonance of
  high-energy cosmic neutrinos},'' 4 2020.

\bibitem{GonzalezGarcia:2008ru}
M.~C. Gonzalez-Garcia and M.~Maltoni, ``{Status of Oscillation plus Decay of
  Atmospheric and Long-Baseline Neutrinos},'' {\em Phys. Lett.}, vol.~B663,
  pp.~405--409, 2008.

\bibitem{Gomes:2014yua}
R.~A. Gomes, A.~L.~G. Gomes, and O.~L.~G. Peres, ``{Constraints on neutrino
  decay lifetime using long-baseline charged and neutral current data},'' {\em
  Phys. Lett.}, vol.~B740, pp.~345--352, 2015.

\bibitem{Choubey:2018cfz}
S.~Choubey, D.~Dutta, and D.~Pramanik, ``{Invisible neutrino decay in the light
  of NOvA and T2K data},'' {\em JHEP}, vol.~08, p.~141, 2018.

\bibitem{Aharmim:2018fme}
B.~Aharmim {\em et~al.}, ``{Constraints on Neutrino Lifetime from the Sudbury
  Neutrino Observatory},'' {\em Phys. Rev.}, vol.~D99, no.~3, p.~032013, 2019.

\bibitem{Lorenz:2021alz}
C.~S. Lorenz, L.~Funcke, M.~L\"offler, and E.~Calabrese, ``{Reconstruction of
  the neutrino mass as a function of redshift},'' {\em Phys. Rev. D}, vol.~104,
  no.~12, p.~123518, 2021.

\bibitem{Escudero:2020ped}
M.~Escudero, J.~Lopez-Pavon, N.~Rius, and S.~Sandner, ``{Relaxing Cosmological
  Neutrino Mass Bounds with Unstable Neutrinos},'' {\em JHEP}, vol.~12, p.~119,
  2020.

\bibitem{Bashinsky:2003tk}
S.~Bashinsky and U.~Seljak, ``{Neutrino perturbations in CMB anisotropy and
  matter clustering},'' {\em Phys. Rev. D}, vol.~69, p.~083002, 2004.

\bibitem{Audren:2014lsa}
B.~Audren {\em et~al.}, ``{Robustness of cosmic neutrino background detection
  in the cosmic microwave background},'' {\em JCAP}, vol.~1503, p.~036, 2015.

\bibitem{Follin:2015hya}
B.~Follin, L.~Knox, M.~Millea, and Z.~Pan, ``{First Detection of the Acoustic
  Oscillation Phase Shift Expected from the Cosmic Neutrino Background},'' {\em
  Phys. Rev. Lett.}, vol.~115, no.~9, p.~091301, 2015.

\bibitem{Baumann:2015rya}
D.~Baumann, D.~Green, J.~Meyers, and B.~Wallisch, ``{Phases of New Physics in
  the CMB},'' {\em JCAP}, vol.~01, p.~007, 2016.

\bibitem{Angrik:2005ep}
J.~Angrik {\em et~al.}, ``{KATRIN design report 2004},'' 2005.

\bibitem{Gerbino:2016ehw}
M.~Gerbino, M.~Lattanzi, O.~Mena, and K.~Freese, ``{A novel approach to
  quantifying the sensitivity of current and future cosmological datasets to
  the neutrino mass ordering through Bayesian hierarchical modeling},'' {\em
  Phys. Lett. B}, vol.~775, pp.~239--250, 2017.

\bibitem{Caldwell:2017mqu}
A.~Caldwell, M.~Ettengruber, A.~Merle, O.~Schulz, and M.~Totzauer, ``{Global
  Bayesian analysis of neutrino mass data},'' {\em Phys. Rev. D}, vol.~96,
  no.~7, p.~073001, 2017.

\bibitem{Vagnozzi:2017ovm}
S.~Vagnozzi, E.~Giusarma, O.~Mena, K.~Freese, M.~Gerbino, S.~Ho, and
  M.~Lattanzi, ``{Unveiling $\nu$ secrets with cosmological data: neutrino
  masses and mass hierarchy},'' {\em Phys. Rev. D}, vol.~96, no.~12, p.~123503,
  2017.

\bibitem{Simpson:2017qvj}
F.~Simpson, R.~Jimenez, C.~Pena-Garay, and L.~Verde, ``{Strong Bayesian
  Evidence for the Normal Neutrino Hierarchy},'' {\em JCAP}, vol.~06, p.~029,
  2017.

\bibitem{DiValentino:2021hoh}
E.~Di~Valentino, S.~Gariazzo, and O.~Mena, ``{Most constraining cosmological
  neutrino mass bounds},'' {\em Phys. Rev. D}, vol.~104, no.~8, p.~083504,
  2021.

\bibitem{Jimenez:2022dkn}
R.~Jimenez, C.~Pena-Garay, K.~Short, F.~Simpson, and L.~Verde, ``{Neutrino
  Masses and Mass Hierarchy: Evidence for the Normal Hierarchy},'' 3 2022.

\bibitem{Schwetz:2017fey}
T.~Schwetz, K.~Freese, M.~Gerbino, E.~Giusarma, S.~Hannestad, M.~Lattanzi,
  O.~Mena, and S.~Vagnozzi, ``{Comment on ''Strong Evidence for the Normal
  Neutrino Hierarchy''},'' 3 2017.

\bibitem{Gariazzo:2018pei}
S.~Gariazzo, M.~Archidiacono, P.~F. de~Salas, O.~Mena, C.~A. Ternes, and
  M.~T\'ortola, ``{Neutrino masses and their ordering: Global Data, Priors and
  Models},'' {\em JCAP}, vol.~03, p.~011, 2018.

\bibitem{Hergt:2021qlh}
L.~T. Hergt, W.~J. Handley, M.~P. Hobson, and A.~N. Lasenby, ``{Bayesian
  evidence for the tensor-to-scalar ratio $r$ and neutrino masses $m_\nu$:
  Effects of uniform vs logarithmic priors},'' {\em Phys. Rev. D}, vol.~103,
  p.~123511, 2021.

\bibitem{Gariazzo:2022ahe}
S.~Gariazzo {\em et~al.}, ``{Neutrino mass and mass ordering: No conclusive
  evidence for normal ordering},'' 5 2022.

\bibitem{Palanque-Delabrouille:2019iyz}
N.~Palanque-Delabrouille, C.~Y\`eche, N.~Sch\"oneberg, J.~Lesgourgues,
  M.~Walther, S.~Chabanier, and E.~Armengaud, ``{Hints, neutrino bounds and WDM
  constraints from SDSS DR14 Lyman-$\alpha$ and Planck full-survey data},''
  {\em JCAP}, vol.~04, p.~038, 2020.

\bibitem{Ma:1994dv}
C.-P. Ma and E.~Bertschinger, ``{Cosmological perturbation theory in the
  synchronous versus conformal Newtonian gauge},'' 1 1994.

\bibitem{Poulin:2016nat}
V.~Poulin, P.~D. Serpico, and J.~Lesgourgues, ``{A fresh look at linear
  cosmological constraints on a decaying dark matter component},'' {\em JCAP},
  vol.~08, p.~036, 2016.

\bibitem{Blinov:2020uvz}
N.~Blinov, C.~Keith, and D.~Hooper, ``{Warm Decaying Dark Matter and the Hubble
  Tension},'' {\em JCAP}, vol.~06, p.~005, 2020.

\bibitem{Audren:2012wb}
B.~Audren, J.~Lesgourgues, K.~Benabed, and S.~Prunet, ``{Conservative
  Constraints on Early Cosmology: an illustration of the Monte Python
  cosmological parameter inference code},'' {\em JCAP}, vol.~02, p.~001, 2013.

\bibitem{Brinckmann:2018cvx}
T.~Brinckmann and J.~Lesgourgues, ``{MontePython 3: boosted MCMC sampler and
  other features},'' {\em Phys. Dark Univ.}, vol.~24, p.~100260, 2019.

\bibitem{Beutler:2011hx}
F.~{Beutler}, C.~{Blake}, M.~{Colless}, D.~H. {Jones}, L.~{Staveley-Smith},
  L.~{Campbell}, Q.~{Parker}, W.~{Saunders}, and F.~{Watson}, ``{The 6dF Galaxy
  Survey: baryon acoustic oscillations and the local Hubble constant},'' {\em
  MNRAS}, vol.~416, pp.~3017--3032, Oct. 2011.

\bibitem{Ross:2014qpa}
A.~J. Ross, L.~Samushia, C.~Howlett, W.~J. Percival, A.~Burden, and M.~Manera,
  ``{The clustering of the SDSS DR7 main Galaxy sample \textendash{} I. A 4 per
  cent distance measure at $z = 0.15$},'' {\em Mon. Not. Roy. Astron. Soc.},
  vol.~449, no.~1, pp.~835--847, 2015.

\bibitem{Alam:2016hwk}
S.~Alam {\em et~al.} {\em Mon. Not. Roy. Astron. Soc.}, vol.~470, no.~3,
  pp.~2617--2652, 2017.

\bibitem{Agathe:2019vsu}
V.~de~Sainte~Agathe {\em et~al.}, ``{Baryon acoustic oscillations at z = 2.34
  from the correlations of Ly$\alpha$ absorption in eBOSS DR14},'' {\em Astron.
  Astrophys.}, vol.~629, p.~A85, 2019.

\bibitem{Blomqvist:2019rah}
M.~Blomqvist {\em et~al.}, ``{Baryon acoustic oscillations from the
  cross-correlation of Ly$\alpha$ absorption and quasars in eBOSS DR14},'' {\em
  Astron. Astrophys.}, vol.~629, p.~A86, 2019.

\bibitem{Scolnic:2017caz}
D.~M. Scolnic {\em et~al.}, ``{The Complete Light-curve Sample of
  Spectroscopically Confirmed SNe Ia from Pan-STARRS1 and Cosmological
  Constraints from the Combined Pantheon Sample},'' {\em Astrophys. J.},
  vol.~859, no.~2, p.~101, 2018.

\bibitem{Gelman:1992zz}
A.~Gelman and D.~B. Rubin {\em Statist. Sci.}, vol.~7, pp.~457--472, 1992.

\bibitem{Archidiacono:2016lnv}
M.~Archidiacono, T.~Brinckmann, J.~Lesgourgues, and V.~Poulin, ``{Physical
  effects involved in the measurements of neutrino masses with future
  cosmological data},'' {\em JCAP}, vol.~02, p.~052, 2017.

\bibitem{Font-Ribera:2013rwa}
A.~Font-Ribera, P.~McDonald, N.~Mostek, B.~A. Reid, H.-J. Seo, and A.~Slosar,
  ``{DESI and other dark energy experiments in the era of neutrino mass
  measurements},'' {\em JCAP}, vol.~05, p.~023, 2014.

\bibitem{Amendola:2016saw}
L.~Amendola {\em et~al.}, ``{Cosmology and fundamental physics with the Euclid
  satellite},'' {\em Living Rev. Rel.}, vol.~21, no.~1, p.~2, 2018.

\bibitem{Liu:2015txa}
A.~Liu, J.~R. Pritchard, R.~Allison, A.~R. Parsons, U.~Seljak, and B.~D.
  Sherwin, ``{Eliminating the optical depth nuisance from the CMB with 21 cm
  cosmology},'' {\em Phys. Rev. D}, vol.~93, no.~4, p.~043013, 2016.

\bibitem{Maartens:2015mra}
R.~Maartens, F.~B. Abdalla, M.~Jarvis, and M.~G. Santos, ``{Overview of
  Cosmology with the SKA},'' {\em PoS}, vol.~AASKA14, p.~016, 2015.

\end{thebibliography}
\bibliographystyle{ieeetr}

\end{document}